\documentclass[sigconf]{acmart}
\usepackage{float}
\usepackage{graphicx}
\usepackage{caption}
\usepackage{booktabs}
\usepackage{tabularx}
\usepackage{hyperref}
\usepackage{makecell}
\usepackage{array}
\AtBeginDocument{%
  }

\copyrightyear{2026}
\acmYear{2026}
\setcopyright{cc}
\setcctype{by}
\acmConference[OzCHI '26]{Proceedings of the 38th Australasian Conference on Human-Computer Interaction}{November 21--25, 2026}{Adelaide, SA, Australia}
\acmBooktitle{Proceedings of the 38th Australasian Conference on Human-Computer Interaction (OzCHI '26), November 21--25, 2026, Adelaide, SA, Australia}

\acmDOI{10.1145/3847401.3847411}

\acmISBN{979-8-4007-2724-5/2026/11}
\begin{document}

\title{How People Use ChatGPT in Australia: A WildChat Analysis}

\author{Ying Ma}
\orcid{0000-0001-5413-0132}
\email{ying.ma@sydney.edu.au}
\affiliation{%
   \department{School of Computer Science}
  \institution{University of Sydney}
  \city{Sydney}
  \country{Australia}}

\author{Katy Gero}
\orcid{0000-0001-5982-9321}
\email{katy.gero@sydney.edu.au}
\affiliation{%
   \department{School of Computer Science}
  \institution{University of Sydney}
  \city{Sydney}
  \country{Australia}}

\author{Clément Canonne}
\orcid{0000-0001-7153-5211}
\email{clement.canonne@sydney.edu.au}
\affiliation{%
   \department{School of Computer Science}
  \institution{University of Sydney}
  \city{Sydney}
  \country{Australia}}

\author{Craig Jin}
\orcid{0000-0003-4636-753X}
\email{craig.jin@sydney.edu.au}
\affiliation{%
   \department{School of Computer Science}
  \institution{University of Sydney}
  \city{Sydney}
  \country{Australia}}

\author{Kanchana Thilakarathna}
\orcid{0000-0003-4332-0082}
\email{kanchana.thilakarathna@sydney.edu.au}
\affiliation{%
   \department{School of Computer Science}
  \institution{University of Sydney}
  \city{Sydney}
  \country{Australia}}









\begin{abstract}
Generative AI chatbots are increasingly embedded in everyday life, yet most large-scale studies describe global patterns. This paper presents an Australia-focused analysis of WildChat, a public dataset of real-world ChatGPT interaction logs. Using descriptive analysis and a multi-layer classification scheme, we analysed 37,845 conversations identified as Australian, examining language diversity, work relevance, interaction intent, topic distribution, turn-taking, temporal change, work activities, and Australia-related domains. Our findings show that the Australian subset is strongly action-oriented and comparatively work-oriented, with most interactions classified as doing and a majority of conversations classified as work-related. The dataset also shows multilingual use and a growing presence of self-expression over time. Australia-related conversations frequently invoke local institutions, laws, regulators, education systems, companies, cultural references, and public services. Finally, we outline implications for future research, including local AI evaluation, multilingual participation, context-aware design, and safeguards for everyday high-stakes domains.
\end{abstract}



%


\maketitle

\section{Introduction}

Generative AI chatbots have rapidly become part of everyday digital life. Systems such as ChatGPT are no longer used only as technical tools or productivity assistants; they are increasingly embedded in diverse activities such as writing, information seeking, learning, practical decision-making, emotional reflection, and creative production. This broadening of use raises an important question for Human–Computer Interaction: how do people actually appropriate general-purpose AI systems in everyday contexts, and how do these patterns differ across social, cultural, and geographic settings?

Recent large-scale studies have begun to answer this question by analysing real-world chatbot interactions. For example, \citet{chatterji2025people} classified ChatGPT messages by work relevance, topic, and user intent, showing that ChatGPT use spans both work and non-work contexts, and includes practical guidance, information seeking, writing, and other everyday activities. WildChat also provides an important public dataset for studying real-world ChatGPT interaction logs at scale, including conversation content, timestamps, model information, language labels, moderation signals, and geographic metadata \cite{zhao2024wildchat}. Together, these studies provide valuable evidence about how people use general-purpose LLM systems beyond controlled laboratory or survey settings.

However, these global accounts leave open how the same broad categories of use are enacted in particular national settings. A conversation labelled as writing, advice-seeking, or information seeking may involve very different forms of knowledge depending on where the user is located. In the Australian context, such requests can be tied to local procedures, institutional expectations, legal frameworks, and public-service pathways.

Australia provides a useful context for examining these issues. It is a multilingual society with English-dominant institutions, a large international student and migrant population, distinctive education and labour-market systems, and ongoing public debates about trustworthy and responsible AI. In this context, ChatGPT may serve multiple roles: a writing and translation assistant, a study and work-support tool, an information-seeking interface, a practical guide for navigating institutions, and a space for personal reflection. Yet we know relatively little about how these roles appear in large-scale interaction data from Australian users.

This paper addresses this gap through an Australia-focused analysis of WildChat. We analyse 37,845 conversations identified as Australia users, spanning from April 2023 to July 2025, and examine how users engage with ChatGPT across multiple dimensions: language diversity, work relevance, interaction intent, topic distribution, turn-taking, temporal change, work activity categories, and Australia-related domains. Rather than treating ChatGPT use as a single topic distribution, we adopt a multi-layered view of interaction. This allows us to examine not only what users ask about, but also whether they use ChatGPT for work or non-work purposes, whether they are asking, doing, or expressing, how conversations vary in length, and how locally grounded Australian references appear in the data.

We ask the following research questions:
\begin{itemize}
    \item RQ1: How is ChatGPT used in the Australian WildChat subset across language, work relevance, interaction intent, and topic, and how do these patterns compare with global patterns reported in prior large-scale studies?
    \item RQ2: How do patterns of ChatGPT use change over time in the Australian subset, particularly across topic prevalence, conversation length, and self-expression?
    \item RQ3: What work activities and Australia-specific domains are represented in the conversations, and what do they reveal about locally situated uses of ChatGPT?

\end{itemize}

Our analysis shows that the Australian subset is strongly action-oriented and comparatively work-oriented.  Following the framework from prior work \cite{chatterji2025people}, most interactions are classified as Doing, and a majority of conversations are classified as work-related. Writing is the largest high-level topic group. 
At the lower-level topic level, translation is the most frequent category.
At the same time, the dataset shows substantial multilingual use and a growing presence of self-expression over time. Australia-related conversations further show that users invoke local institutions, laws, regulators, education systems, companies, cultural references, and public services.

In this paper, we provide an empirical account of ChatGPT use in an Australian subset of WildChat, showing how usage patterns differ from broader global trends reported in prior work. We also outline implications for future HCI research and design based on the patterns and trends observed in the Australian subset. Our findings highlight the need for locally grounded AI evaluation, better support for multilingual communication, context-aware design for institution-facing tasks, and safeguards for everyday high-stakes domains where users may rely on AI-generated guidance.


\section{Related Work}
This section reviews prior work that informs our analysis of ChatGPT use in Australia. We begin with large-scale studies of real-world LLM interactions, which provide a basis for examining work relevance, topics, and user intent. We then discuss research on generative AI in work and everyday task support, showing why productivity-oriented use extends beyond formal employment. Finally, we introduce work on locally situated chatbot use.

\subsection{Large-Scale Evidence of LLM Use}

As generative AI systems have moved from specialist tools into everyday use, HCI research increasingly needs empirical accounts of how people actually interact with them at scale. Early chatbot research often focused on conversational agents as designed systems for specific domains, such as customer service, education, health, or companionship \citep{adamopoulou2020overview}. However, general-purpose LLM-based systems such as ChatGPT differ from earlier chatbots because they are open-ended: the same system can be used for writing, programming, translation, information seeking, advice, emotional reflection, and creative production. This makes it difficult to understand use through predefined application domains alone.

Recent large-scale studies have begun to address this challenge by analysing real-world interaction traces. WildChat provides one important resource for this kind of work, releasing a large corpus of ChatGPT conversations collected in the wild and including metadata such as timestamps, language, model, and geographic signals \citep{zhao2024wildchat}. Such datasets make it possible to study not only what users ask LLMs to do, but also how use varies across contexts and over time. Complementing this dataset-level contribution, \citet{chatterji2025people} analyse ChatGPT use through categories such as work relevance, topic, and user intent. Their work shows that ChatGPT use is not limited to technical or professional tasks; instead, it spans practical guidance, information seeking, writing, and other everyday activities. Their findings also suggest that non-work use has become increasingly prominent in global ChatGPT activity.

\begin{figure*}[t]
\centering 
\includegraphics[width=1\textwidth]{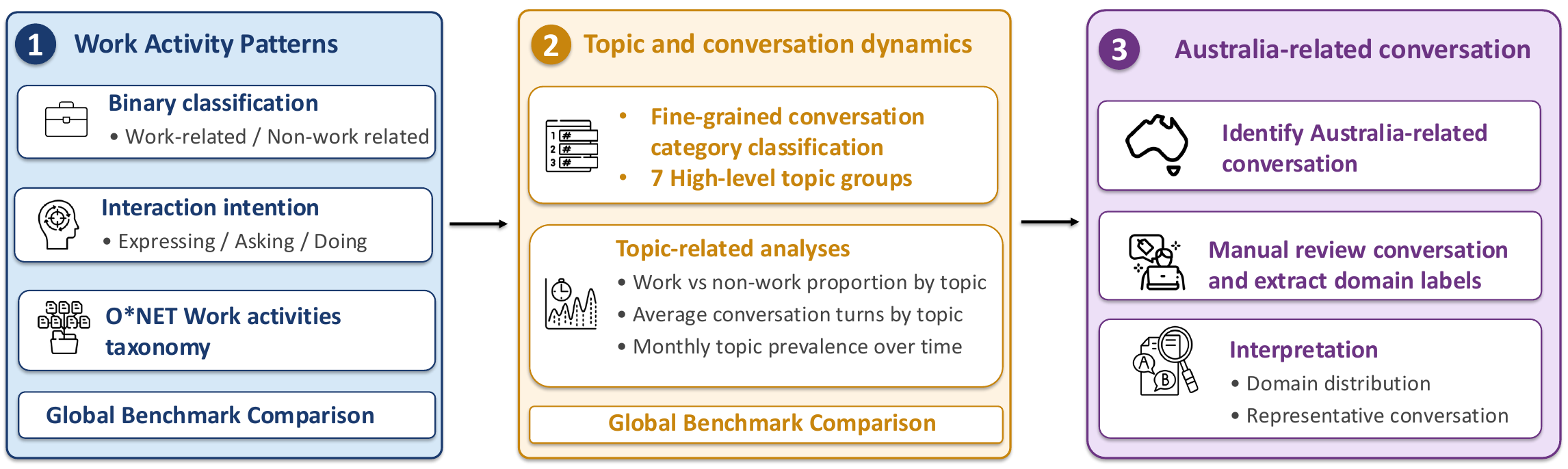} 
\caption{Overview of the study methodology.} 
\Description{A three-stage overview of the study methodology. Stage 1 classifies conversations by work relevance, interaction intent, and O*NET work activities, followed by comparison with global benchmarks. Stage 2 classifies conversation topics and analyses topic-level work relevance, conversation length, and monthly prevalence. Stage 3 identifies Australia-related conversations, manually develops domain labels, and examines their distribution and representative examples.}
\label{fig:method} 
\end{figure*}

These studies provide a valuable foundation for understanding general patterns of LLM use. However, their broad scope also raises an important HCI question: whether global patterns adequately describe local or national contexts. Aggregated platform-level analyses can show dominant use categories, but they may obscure how people in particular regions appropriate LLMs in relation to local languages, institutions, work practices, and everyday concerns. This is especially relevant for countries such as Australia, where public discussions of AI are shaped by multilingual communities, work and education contexts, and emerging debates around trustworthy and responsible AI. Our work builds on large-scale studies of ChatGPT use, but shifts the unit of analysis toward an Australian subset of WildChat. In doing so, we examine whether Australian conversations mirror global patterns or reveal a more locally specific profile of generative AI use.

\subsection{Generative AI for Work and Everyday Task Support}

A major strand of prior work has examined generative AI as a tool for productivity and labour. Experimental evidence suggests that generative AI can improve performance on writing-intensive professional tasks, particularly by helping users produce, revise, and structure text more efficiently \citep{noy2023experimental}. Other studies examine the broader labour-market implications of LLMs, arguing that many occupations contain tasks that may be affected by language-model capabilities \citep{eloundou2024gpts}. In workplace settings, generative AI has also been studied as a support tool for customer service and other knowledge-work environments, where it can affect worker productivity and task completion \citep{brynjolfsson2025generative,zhang2026investigating}. Together, this work positions LLMs as increasingly important infrastructures for work, productivity, and professional communication.

At the same time, HCI research needs to avoid treating ``work use'' too narrowly. Many LLM-supported activities are task-oriented without being confined to formal employment. Users may ask for help writing an email, summarising information, learning a concept, preparing an application, debugging code, or planning a practical decision. Some of these activities may be professional, some educational, and some personal, but they share a common structure: the user approaches the system with a concrete goal and expects the model to produce, transform, or evaluate an output. This distinction is important because a simple work/non-work binary can hide the broader role of LLMs as everyday task-support systems.

The taxonomy used by \citet{chatterji2025people} is useful because it separates work relevance from user intent and topic. For example, writing can occur in both work and non-work contexts; seeking information may support professional research, study, or personal decision-making; and technical help may include both occupational programming and informal learning. Our analysis follows this multi-layered view. Rather than asking only whether Australian conversations are work-related, we examine how work relevance intersects with interaction intent, topic, turn-taking, sentiment, and work activity categories. This approach helps show whether Australian use is best understood as workplace productivity, broader task orientation, or a mixture of professional, educational, practical, and personal goals.

\subsection{Affective, Cultural, and Locally Situated Chatbot Use}

Although productivity and task support are central to many accounts of generative AI, prior chatbot research also shows that conversational agents can become spaces for personal disclosure, emotional expression, and social interaction. Studies of chatbot companionship suggest that users may form meaningful relationships with conversational agents and use them for support, reflection, and ongoing interaction \citep{skjuve2021my}. Related work on digital confession and self-disclosure shows that people may be willing to share intimate information with chatbots, with potential implications for emotional wellbeing and the design of conversational systems \citep{croes2024digital}. These findings highlight that chatbot use is not only instrumental: it can also be affective, relational, and personally meaningful \cite{brandtzaeg2017people}.


At the same time, recent national and regional LLM initiatives suggest that language models are increasingly understood as locally situated infrastructure rather than only as general-purpose global systems. Projects such as OpenEuroLLM aim to develop transparent and multilingual foundation models for European languages, motivated by concerns around linguistic diversity, regulatory alignment, and digital sovereignty \citep{openeurollm2025}. Similarly, LLM-jp was established as a collaborative hub for Japanese LLM research and development, bringing together academic, industry, and infrastructure support to advance open innovation in Japan \citep{aizawa2024llm}. Other initiatives, such as Polish LLM projects including Bielik and PLLuM, emphasise the importance of models trained on local language data for education, business, public administration, and the protection of national and cultural identity \citep{cyfronet2026polishllms}. These examples show that LLM development and evaluation are shaped by local languages, institutions, legal expectations, and cultural contexts. 

Relatedly, prior work has shown that generative AI models often encode culturally uneven assumptions, including Western- or US-centric norms, values, and communicative styles. For example, studies of cultural bias in generative AI show that models may fail to represent non-Western cultural artifacts, reproduce dominant cultural defaults, and homogenise writing toward Western norms \cite{agarwal2025ai}. Recent benchmarks such as CulturalBench and NormAd further show that LLMs continue to struggle with multicultural knowledge and socio-cultural reasoning across diverse regions and norms \cite{chiu2024culturalbench,rao2025normad}. These findings reinforce the need to study ChatGPT use not only as a global platform-level phenomenon, but also as a locally situated interactional practice shaped by users’ cultural, linguistic, and institutional contexts.



\section{Method}

\begin{table*}[t]
\centering
\caption{Topic groups and conversation categories used in the analysis \cite{chatterji2025people}.}
\label{tab:topic_categories}
\small
\setlength{\tabcolsep}{4pt}
\renewcommand{\arraystretch}{1}
\begin{tabularx}{\textwidth}{p{0.2\textwidth} X}
\toprule
\textbf{Topic} & \textbf{Conversation Category} \\
\midrule

Writing &
Edit or Critique Provided Text; Personal Writing or Communication; Translation; Argument or Summary Generation; Write Fiction. \\

\midrule

Practical Guidance &
How-To Advice; Tutoring or Teaching; Creative Ideation; Health, Fitness, Beauty, or Self-Care. \\

\midrule

Technical Help &
Mathematical Calculation; Data Analysis; Computer Programming. \\

\midrule

Multimedia &
Create an Image; Analyze an Image; Generate or Retrieve Other Media. \\

\midrule

Seeking Information &
Specific Info; Purchasable Products; Cooking and Recipes. \\

\midrule

Self-Expression &
Greetings and Chitchat; Relationships and Personal Reflection; Games and Role Play. \\

\midrule

Other/Unknown &
Asking About the Model; Other; Unclear. \\

\bottomrule
\end{tabularx}
\end{table*}

\subsection{Dataset and Scope}
We used WildChat-4.8M, a public dataset of real-world ChatGPT interaction logs \cite{zhao2024wildchat}. It was created by offering online users free access to chatbot services in exchange for their affirmative, consensual opt-in to anonymously collect their chat transcripts and request headers. The extended WildChat-4.8M release covers conversations collected from April 9, 2023 to July 31, 2025.
WildChat dataset contains full user \& assistant conversations and rich metadata for each conversation. This metadata includes the model used, timestamp, number of turns, detected conversation language, moderation results, toxicity and redaction indicators, and request-level information such as country, state, hashed IP address, and request header. 
For this study, we focused on conversations identified as related to Australia (conversations with Australia as the metadata country). The resulting dataset contains 37,845 conversations.

The dataset also includes several safeguards relevant to ethical analysis, including moderation labels, toxicity indicators, and redaction flags. The released non-toxic version excludes conversations flagged as toxic, and personal or sensitive information was de-identified where detected. Nevertheless, we treat the dataset as sensitive interaction data and report only aggregate patterns rather than identifying users or reproducing private conversational details.

\subsection{Data analysis}


Following prior large-scale analyses of ChatGPT use \cite{chatterji2025people}, we adopted a multi-layer classification scheme to characterise how users in the Australian WildChat subset interacted with ChatGPT. 
Figure \ref{fig:method} provides an overview of the study methodology.
We first conducted descriptive analysis using the metadata provided by WildChat, including the number of conversations, collection period, detected languages, and language distribution. This provided an overview of the scale and linguistic diversity of the Australian subset. We then analysed the conversations across four main dimensions: work relevance and work activities, topic and interaction patterns, comparative patterns across global and country-level benchmarks \footnote{https://openai.com/signals/data/}, and Australia-related local contexts.

\subsubsection{Work Relevance and Work Activity Classification}
\label{app:Work_Relevance_and_Work}
First, we classified each user request as either work-related or non-work-related. Work-related requests were defined as requests that appeared to involve professional, academic, organisational, technical, or productivity-oriented activity. Non-work requests included personal, recreational, reflective, or everyday-life uses. This classifier followed the binary format described in Appendix~\ref{app:work-nonwork}.

Then we classified the interactional intent of each request using three categories: \textit{Expressing}, \textit{Asking}, and \textit{Doing}. \textit{Expressing} captures cases where the user primarily shares feelings, opinions, self-reflections, or conversational statements without clearly requesting information or task completion. \textit{Asking} captures cases where the user seeks information, advice, explanation, or evaluation to support understanding or decision-making. \textit{Doing} captures cases where the user asks ChatGPT to perform or produce an output, such as writing, editing, translating, coding, summarising, analysing data, or generating media. The full definitions and examples used to distinguish these three categories are reported in Appendix~\ref{app:expressing-asking-doing}.

For conversations classified as work-related, we employed the O*NET Work Activities taxonomy as a structured framework to categorize the types of work activities for which users sought assistance from ChatGPT, following methodologies adopted in prior studies \cite{tomlinson2025working,chatterji2025people}. O*NET’s work-activity hierarchy includes multiple levels of granularity, including Generalized Work Activities (GWAs), Intermediate Work Activities (IWAs), and Detailed Work Activities (DWAs) as a structured taxonomy for describing the types of work users asked ChatGPT to support \footnote{https://www.onetcenter.org/dl\_files/database/db\_30\_3\_text/GWAs\%20to\%20IWAs.txt}.

\subsubsection{Topic and Conversation Dynamics}

Next, we classified each request into a fine-grained conversation category and then aggregated these categories into seven higher-level topic groups: \textit{Writing}, \textit{Practical Guidance}, \textit{Technical Help}, \textit{Multimedia}, \textit{Seeking Information}, \textit{Self-Expression}, and \textit{Other/Unknown}. The fine-grained categories were based on the capability labels defined in the conversation topic classifier, including categories such as 
edit or critique provided text, translation, computer programming, data analysis, and relationships and personal reflection. Table~\ref{tab:topic_categories} shows how these fine-grained categories were grouped for analysis.
The full classifier prompt, including the complete list of labels, is provided in Appendix~\ref{app:conversation-topic}.

We then examined how topic categories related to other dimensions of use. To analyse the relationship between topic and work relevance, we calculated the proportion of work-related and non-work-related conversations within each topic group. To examine conversation depth, we calculated the average number of turns for each low-level topic, excluding invalid labels and topics with insufficient conversation counts to avoid unstable estimates. Finally, we analysed temporal change by calculating monthly topic prevalence across the collection period.

To contextualise patterns observed in the Australian WildChat subset, we conducted two forms of comparative analysis. First, we compared the Australian subset with the OpenAI Signals global comparison reported in prior work~\cite{chatterji2025people}. This comparison focused on dimensions that were defined consistently across the two analyses, including work-related versus non-work-related use, interaction intent, high-level topic groups, and topic composition within work and non-work conversations. For each dimension, we calculated percentage-point differences between the Australian subset and the global comparison.

\subsubsection{Australia-Related Context and Domain Coding}

Finally, we analysed conversations that explicitly referred to Australian contexts. We first identified Australia-related conversations using a classifier prompt reported in Appendix \ref{app:australia-context}. The classifier captured conversations with explicit Australian geographic, institutional, legal, cultural, educational, commercial, or public-service contexts, and was instructed to be conservative when evidence of Australian relevance was ambiguous. We then manually reviewed 100 Australia-related conversations to inductively develop the domain coding scheme. This process produced ten domain labels: culture/media, work/employment, consumer services, education, legal/government, travel/mobility, housing, healthcare, immigration, and other. We applied these labels to the full Australia-related subset using the domain-classification prompt reported in Appendix \ref{app:australia-domain}. 
A comparison between the AI and human annotations of 100 randomly selected conversations showed substantial agreement (Cohen’s $\kappa$ = 0.715).
We then examined both the distribution of domains and representative conversations within each domain to interpret how Australian context was embedded in users’ interactions with ChatGPT.


\subsection{Limitations}

This method has several limitations.
First, WildChat does not represent all ChatGPT use. It captures only conversations available through the dataset collection process and may overrepresent certain users, platforms, models, or usage contexts. Therefore, patterns observed in this dataset may be shaped by dataset composition, user self-selection, model availability, and collection period.

Second, our analysis relies on LLM-assisted labelling, which introduces uncertainty. For work relevance, interaction intent, and topic classification, we adopted classifier prompts and category definitions from prior large-scale work on ChatGPT use, where these classification schemes were developed and validated. For this reason, we did not conduct a separate manual validation study or report inter-rater reliability for these reused classifiers. Although structured prompts and validity checks help improve consistency, labels such as work relevance, user intent, and topic group still involve interpretation. Some conversations contain multiple goals, limited context, or ambiguous boundaries between work, study, personal communication, and everyday task support, meaning that assigning a dominant label may simplify more complex forms of use.

Third, the comparative analyses should be interpreted descriptively rather than as evidence of nationally representative differences in ChatGPT use. The Australian WildChat subset, the additional country-level WildChat samples, and the OpenAI Signals global comparison were not produced through a single controlled sampling design and may differ in data source, collection period, user composition, and classification procedures. Although these comparisons provide useful context for interpreting the Australian subset, they should not be treated as controlled cross-national comparisons.

\section{Results}

\subsection{Data Overview and Language Diversity}

\begin{figure}[t] \centering \includegraphics[width=0.5\textwidth]{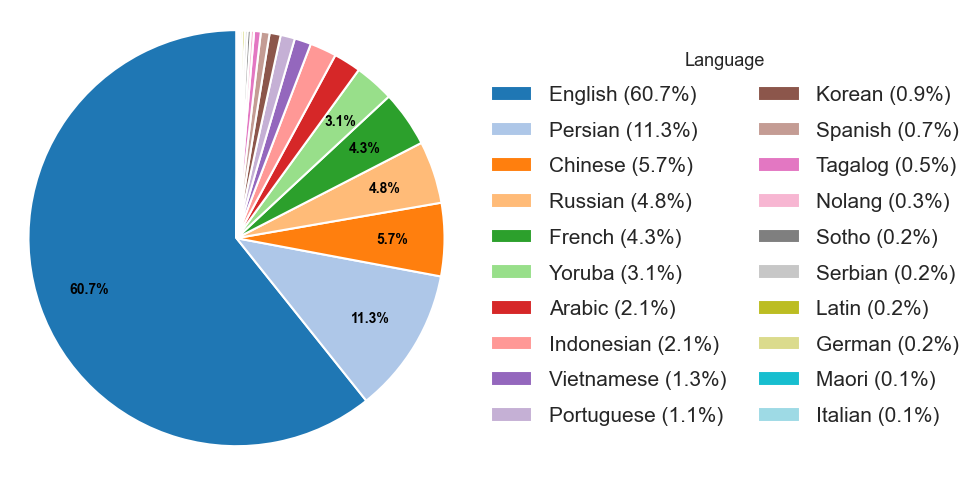} \caption{Language distribution in the Australian WildChat subset.} 
\Description{A pie chart showing the language distribution of conversations in the Australian WildChat subset. English accounts for 60.7 percent, followed by Persian at 11.3 percent, Chinese at 5.7 percent, Russian at 4.8 percent, French at 4.3 percent, Yoruba at 3.1 percent, and a long tail of other languages.}
\label{fig:language_distribution} \end{figure}

The Australian subset contains 37,845 conversations spanning 65 detected languages, suggesting that ChatGPT use in Australia is linguistically diverse rather than limited to English-only interaction. Figure~\ref{fig:language_distribution} shows the distribution across languages. 
English dominates the dataset, accounting for 60.7\% of conversations. However, a substantial proportion of conversations occur in other languages, including Persian (11.3\%), Chinese (5.7\%), Russian (4.8\%), French (4.3\%), and Yoruba (3.1\%), followed by a long tail of additional languages appearing in smaller proportions.
This distribution indicates that ChatGPT is used in Australia across a broad multilingual context. The presence of many non-English conversations suggests that users may rely on ChatGPT for multilingual writing, translation, communication, and information access.

\subsection{Work Relevance and Work Activity Patterns}

\begin{figure}[H]
    \centering

    \begin{minipage}[t]{0.48\textwidth}
        \centering
        \includegraphics[width=\textwidth]{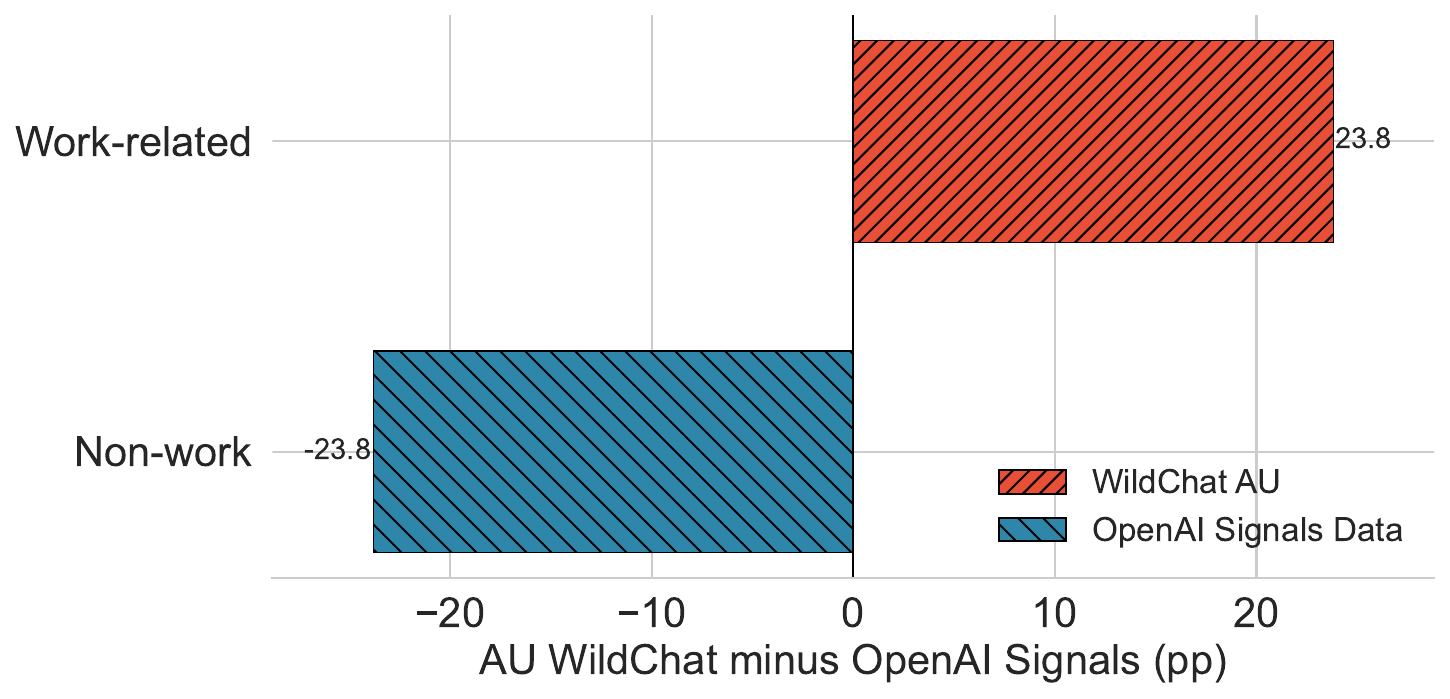}
        \caption{Work/non-work gap (AU vs global dataset).}
        \Description{A diverging horizontal bar chart comparing work and non-work conversation shares in the Australian WildChat subset with the OpenAI Signals global data. The Australian subset has a 23.8 percentage-point higher share of work-related conversations and a 23.8 percentage-point lower share of non-work conversations.}
        \label{fig:worksignalgap}
    \end{minipage}
    \hfill
    \begin{minipage}[t]{0.48\textwidth}
        \centering
        \includegraphics[width=\textwidth]{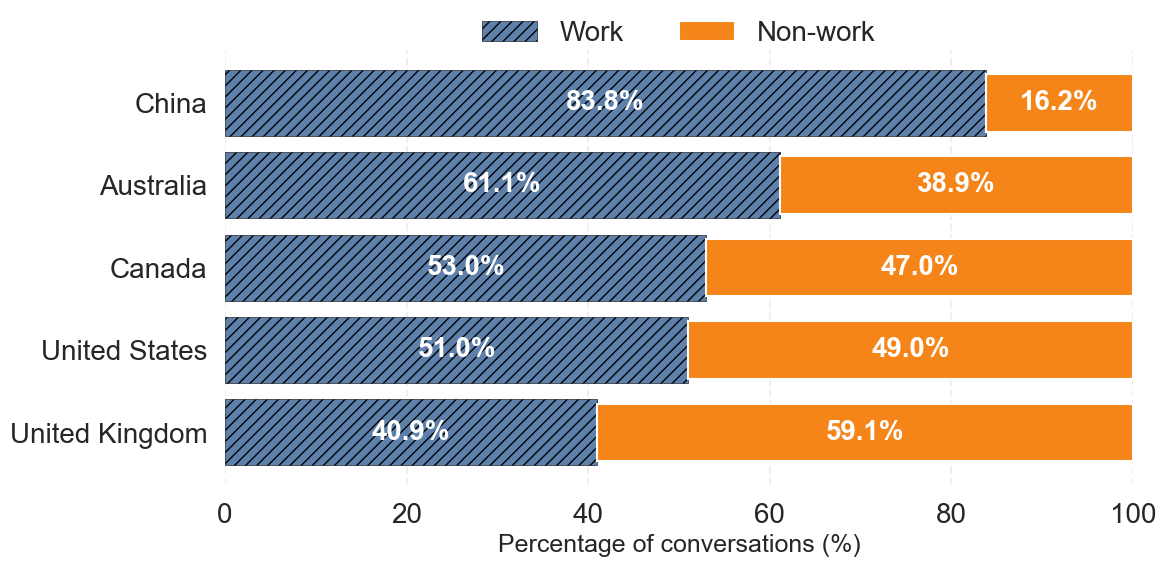}
        \caption{Work and Non-Work Conversation Distribution Across Countries.}
        \Description{A horizontal bar chart comparing the proportions of work-related and non-work conversations across five country groups. China has the highest work-related share at 83.8 percent, followed by Australia at 61.1 percent, Canada at 53.0 percent, the United States at 51.0 percent, and the United Kingdom at 40.9 percent.}
        \label{fig:workaccrosscountry}
    \end{minipage}

\end{figure}

\textbf{Work and Non-Work Use}: 
The Australian WildChat subset is strongly work-oriented. Work-related conversations account for 61.1\% of the Australian sample, while non-work conversations account for 38.9\%. In contrast, the OpenAI Signals global comparison shows a lower share of work-related use (37.3\%) and a higher share of non-work use (62.7\%). As shown in Figure~\ref{fig:worksignalgap}, this corresponds to a 23.8 percentage-point higher share of work-related conversations in the Australian WildChat subset and a 23.8 percentage-point lower share of non-work conversations. 

To place this result in a cross-country context, we randomly sampled 8,000 conversations from each of four additional country groups and labelled them using the same work/non-work classification procedure. Figure~\ref{fig:workaccrosscountry} shows that Australia has a higher proportion of work-related conversations than Canada (53.0\%), the United States (51.0\%), and the United Kingdom (40.9\%), but a lower proportion than China (83.8\%). Together, these comparisons suggest that the Australian subset is comparatively work-oriented, both relative to the global OpenAI Signals benchmark and to several other English-speaking country groups.


\begin{figure}[H]
       \centering
        \includegraphics[width=0.5\textwidth]{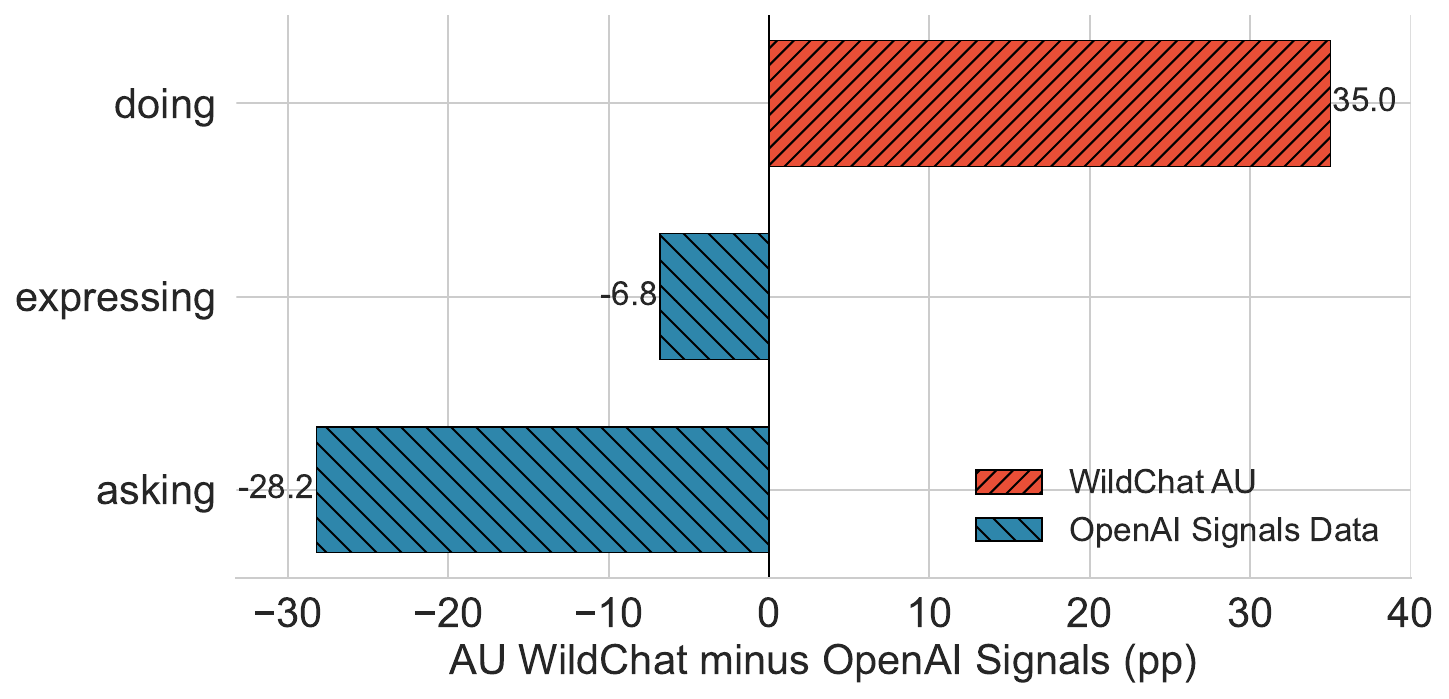}
        \caption{Intention gap (AU vs global dataset).}
        \Description{A diverging horizontal bar chart comparing interaction-intent distributions in the Australian WildChat subset with the OpenAI Signals global data. Doing is 35.0 percentage points higher in the Australian subset, while Expressing is 6.8 points lower and Asking is 28.2 points lower.}
        \label{fig:intentiongap}
\end{figure}

\textbf{Interaction Intent}: We categorize user interactions into three intent labels, \emph{Expressing}, \emph{Asking}, and \emph{Doing}. The Australian sample is strongly action-oriented: \emph{Doing} accounts for 74.8\% of interactions, compared with 39.8\% in the OpenAI Signals global comparison. As shown in Figure~\ref{fig:intentiongap}, this is the largest gap across the three intent categories, with \emph{Doing} 35.0 percentage points higher in the Australian subset. By contrast, \emph{Asking} is substantially lower in Australia (13.9\% vs. 42.1\%, a gap of -28.2 percentage points), while \emph{Expressing} is also lower but to a smaller extent (11.3\% vs. 18.1\%, a gap of -6.8 percentage points). These results suggest that Australian users in this dataset engage ChatGPT more as a task-execution tool than as an information-seeking or expressive conversational system.

\subsubsection{\textbf{O*NET Work Activities}:}
\begin{figure}[H]
    \centering
    \includegraphics[width=0.5\textwidth]{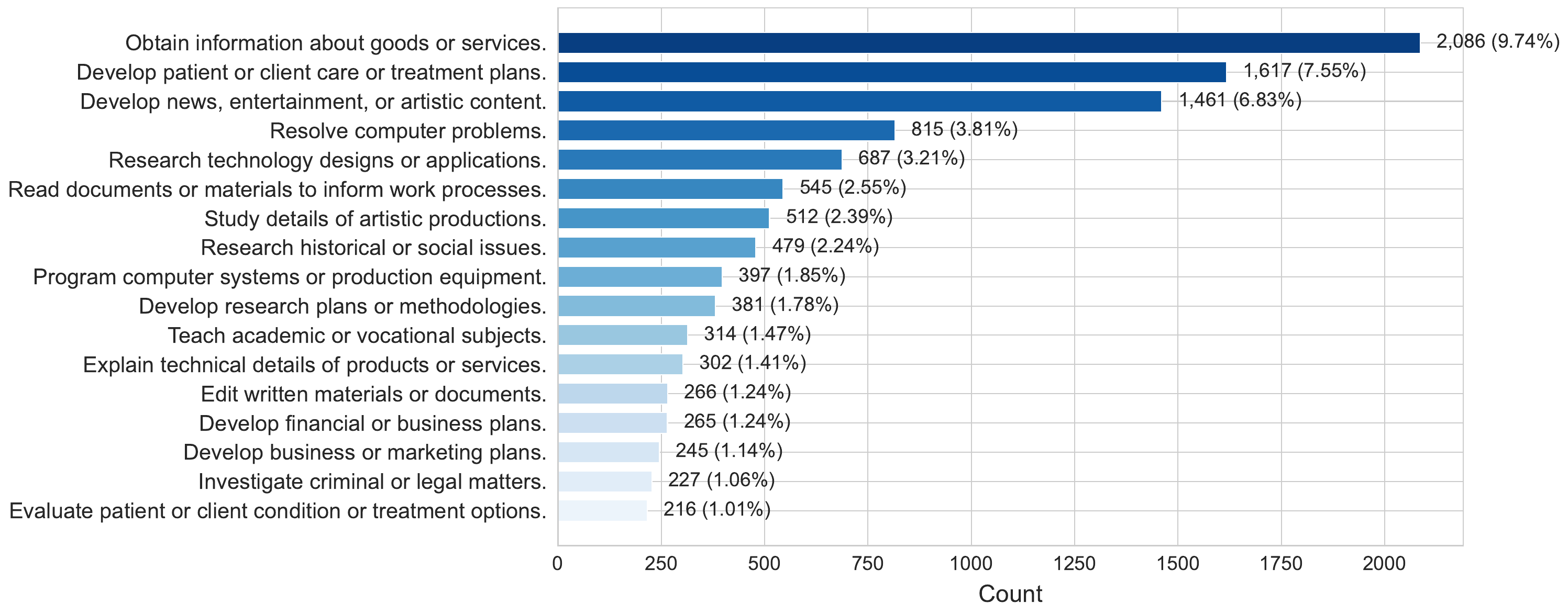}
    \caption{Distribution of the 17 most frequent O*NET Intermediate Work Activities among work-related conversations.}
    \Description{A ranked horizontal bar chart showing the 17 most frequent named O*NET Intermediate Work Activities among work-related conversations. Obtaining information about goods or services is the most frequent activity with 2,086 conversations, followed by developing patient or client care or treatment plans with 1,617 and developing news, entertainment, or artistic content with 1,461.}
    \label{fig:ONET}
\end{figure}

To understand the work activities associated with ChatGPT usage, we mapped messages to one of the 332 O*NET Intermediate Work Activities (IWA) \footnote{https://www.onetcenter.org/dl\_files/database/db\_20\_1\_text/IWA\%20Reference.txt}.
There are 21,406 work-related conversations received a non-missing IWA label; the remainder reflect parsing or API failures. The distribution is highly concentrated but still spans many activity types: the largest single category is “unclear” (4,008 conversations; 18.7\% of labelled cases), followed by information gathering and knowledge work—e.g., obtaining information about goods or services (9.7\%), developing care or treatment plans (7.6\%), developing news or entertainment content (6.8\%), and resolving computer problems (3.8\%). Figure \ref{fig:ONET} summarizes the 17 most frequent named IWAs (ranks 2–18), which together account for 10,815 labelled conversations (50.5\% of all valid assignments, or 62.2\% of assignments excluding the unclear category). Beyond this head, activity types tail off quickly: 249 distinct IWA codes appear at least once, including research, programming, teaching, editing, planning, and professional advising. These patterns suggest that open-ended chatbot use in work contexts clusters around retrieval, content production, technical troubleshooting, and professional planning, while a non-trivial share of exchanges resist alignment with a single standardized work activity. The frequency table for the top 80 IWA categories, including counts and percentages, is reported in Appendix \ref{tab:onenote}.

\subsection{Topic Structure and Conversation Dynamics}

\begin{figure}[H]
        \centering
        \includegraphics[width=0.5\textwidth]{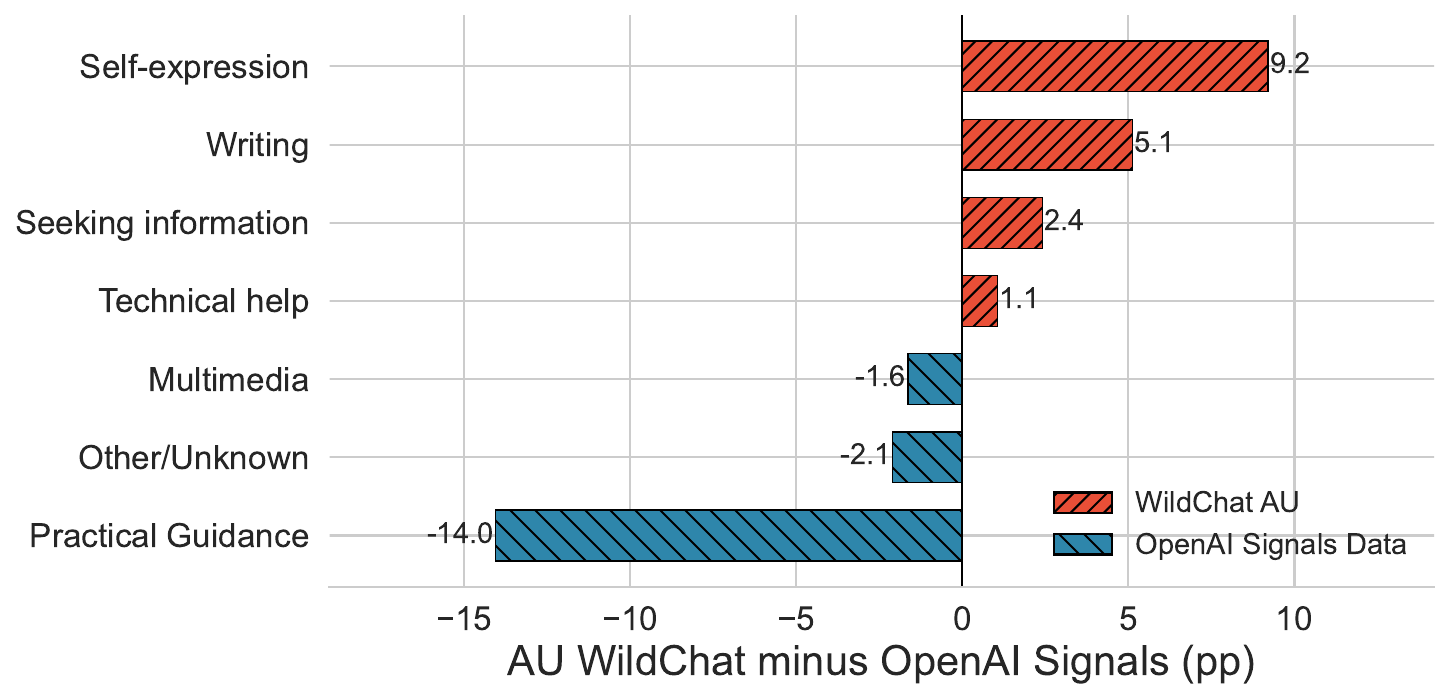}
        \caption{Topic gap (AU vs global dataset).}
        \Description{A diverging horizontal bar chart comparing high-level topic shares in the Australian WildChat subset with the OpenAI Signals global data. Self-Expression is 9.2 percentage points higher in the Australian subset, Writing is 5.1 points higher, Seeking Information is 2.4 points higher, and Technical Help is 1.1 points higher. Multimedia, Other or Unknown, and Practical Guidance are lower by 1.6, 2.1, and 14.0 percentage points, respectively.}
        \label{fig:topicgap}
\end{figure}

\begin{figure*}[t]
    \centering
    \includegraphics[width=\textwidth]{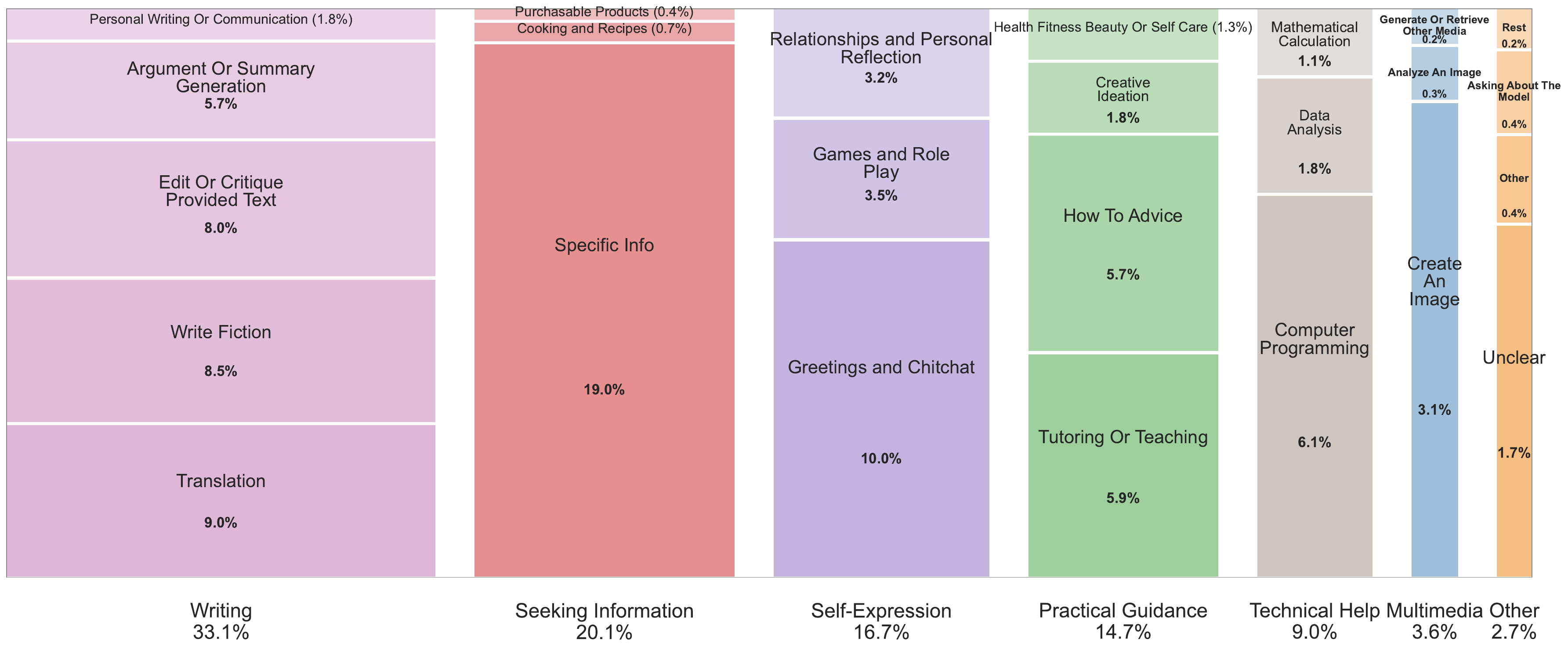}
    \caption{Distribution of grouped high-level topics.}
    \Description{This figrue showing the distribution of low-level conversation topics, grouped into seven higher-level categories. The largest individual topics include Specific Information at 19.0 percent, Greetings and Chitchat at 10.0 percent, Translation at 9.0 percent, Write Fiction at 8.5 percent, and Edit or Critique Provided Text at 8.0 percent. Smaller areas represent technical help, practical guidance, multimedia, and other topics.}
    \label{fig:high_level_topics}
\end{figure*}

\textbf{Topic Distribution}: Topic analysis shows that Australian ChatGPT conversations are concentrated in several high-frequency categories, as shown in Figure~\ref{fig:high_level_topics}. Writing is the largest topic group, accounting for 33.17\% of conversations, followed by Seeking Information (20.17\%), Self-Expression (16.73\%), Practical Guidance (14.73\%), and Technical Help (8.97\%). 
Compared with the OpenAI Signals global comparison, the Australian subset is less dominated by Practical Guidance and more concentrated in Writing and Self-Expression. Practical Guidance accounts for 14.7\% of Australian conversations, compared with 28.8\% globally, a gap of -14.0 percentage points. By contrast, Self-Expression is more prominent in the Australian subset (16.7\% vs. 7.5\%, +9.2 percentage points), followed by Writing (33.2\% vs. 28.1\%, +5.1 percentage points). Seeking Information and Technical Help are only slightly higher in the Australian subset, while Multimedia and Other/Unknown are lower. As shown in Figure~\ref{fig:topicgap}, these differences suggest that the Australian subset is characterised by comparatively less practical-guidance use and more writing- and self-expression-oriented interaction.

At the lower-level capability layer, the most frequent labels are \texttt{translation} (9.06\%), \texttt{write fiction} (8.48\%), \texttt{edit or critique provided text} (8.07\%), \texttt{computer programming} (6.07\%), \texttt{tutoring or teaching} (5.85\%), and \texttt{how to advice} (5.68\%). These lower-level categories show that writing-related activities include both creative and practical forms of text production, while technical and educational uses also account for a notable share of conversations.

\subsubsection{\textbf{Work-Related Use Varies by Topic}:}
Work-related use varies substantially across topic groups.
Technical Help is the most work-oriented category, with 96.64\% conversations classified as work-related, followed by Multimedia (80.81\%), Practical Guidance (79.56\%), Writing (70.55\%), and Seeking Information (64.13\%). By contrast, Self-Expression is predominantly non-work-related, with 94.70\% of conversations classified as non-work.

\begin{figure*}[t]
    \centering
    \includegraphics[width=\textwidth]{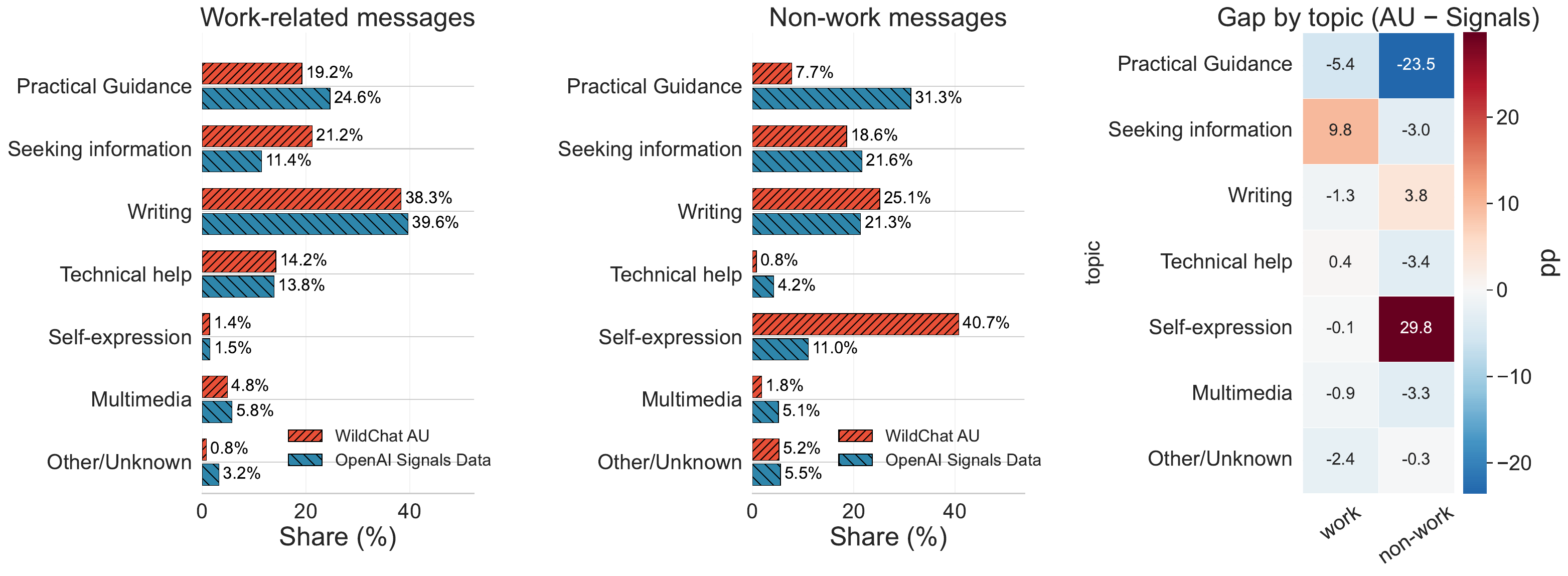}
    \caption{Topic composition of work and non-work conversations (AU vs global dataset).}
    \Description{A three-panel comparison of topic composition in work-related and non-work conversations from the Australian WildChat subset and OpenAI Signals data. Writing is the largest work-related topic in both datasets. Australian work-related conversations contain more Seeking Information and less Practical Guidance. Australian non-work conversations contain substantially more Self-Expression and less Practical Guidance than the global comparison.}
    \label{fig:topic_work_sidebyside}
\end{figure*}

Figure~\ref{fig:topic_work_sidebyside} further compares the topic composition of work and non-work conversations between the Australian WildChat subset and the OpenAI Signals global comparison. Among work-related conversations, Writing is the largest topic in both datasets, accounting for 38.32\% of Australian work-related conversations and 39.64\% of Signals work-related conversations. However, Seeking Information is more prominent in the Australian work subset (21.18\% vs. 11.42\%, +9.75 percentage points), while Practical Guidance is lower (19.20\% vs. 24.63\%, -5.43 percentage points). Technical Help is similar across the two datasets (14.20\% vs. 13.82\%). 

The contrast is stronger among non-work conversations. Self-Expression accounts for 40.73\% of Australian non-work conversations, compared with 10.98\% in the Signals comparison, a gap of +29.76 percentage points. Conversely, Practical Guidance is much less common in Australian non-work conversations (7.74\% vs. 31.27\%, -23.53 percentage points). These patterns suggest that the Australian subset is not only more work-oriented overall, but also differs in the kinds of topics that appear within work and non-work use: work-related conversations are more information-seeking, while non-work conversations are especially concentrated in self-expression.



\subsubsection{\textbf{Average Conversation Turn Number by Low-level Topic}:} 
\begin{figure}[H]
\centering
\includegraphics[width=0.5\textwidth]{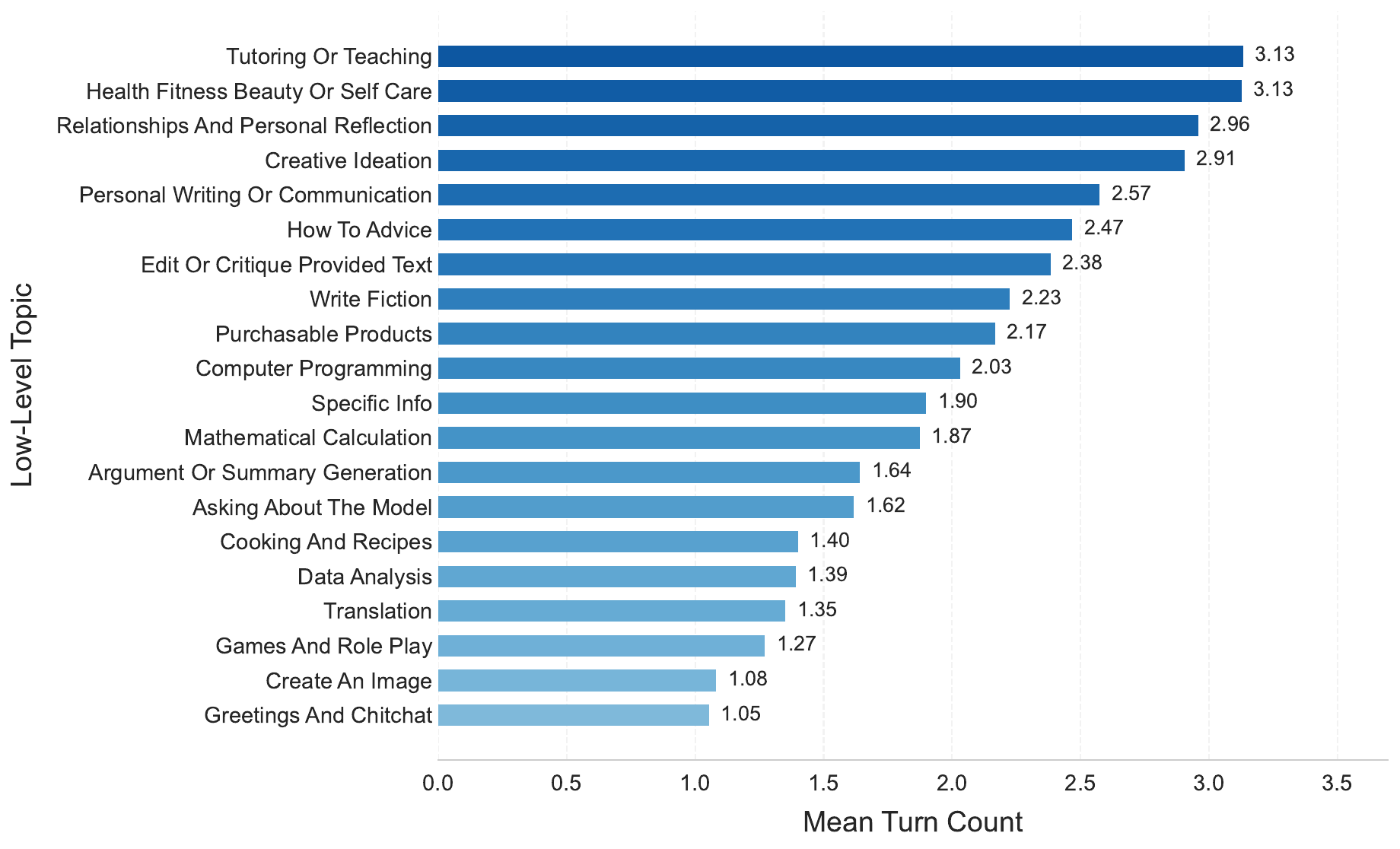}
\caption{Average conversation turn number by low-level topic.}
\Description{A horizontal bar chart showing the mean number of conversation turns for the 20 most frequent eligible low-level topics. Tutoring or Teaching and Health, Fitness, Beauty, or Self-Care have the highest mean at 3.13 turns, followed by Relationships and Personal Reflection at 2.96 and Creative Ideation at 2.91. Greetings and Chitchat, image creation, translation, and data analysis have shorter average conversations.}
\label{fig:turn_level_topics}
\end{figure}

We also examined conversation length across low-level topics by calculating the mean number of turns per topic. To avoid unstable estimates from sparse or noisy labels, we excluded invalid labels and only visualized topics with at least 100 conversations. Figure~\ref{fig:turn_level_topics} shows the 20 most frequent topics among those meeting this threshold. The results suggest that longer conversations are associated with topics requiring iterative clarification or refinement. Tutoring or Teaching and Health, Fitness, Beauty, or Self-care have the highest mean turn counts (M = 3.13), followed by Relationships and Personal Reflection (M = 2.96), Creative Ideation (M = 2.91), and Personal Writing or Communication (M = 2.57). In contrast, more transactional topics such as Greetings and Chitchat (M = 1.05), Create an Image (M = 1.08), Translation (M = 1.35), and Data Analysis (M = 1.39) tend to involve shorter exchanges. This indicates that conversation length varies by task type, with learning, advice, reflection, and ideation generating more sustained interaction than single-output tasks.




\subsubsection{\textbf{Topic Dynamics Over Time}:}
\begin{figure}[t]
\centering
\includegraphics[width=0.5\textwidth]{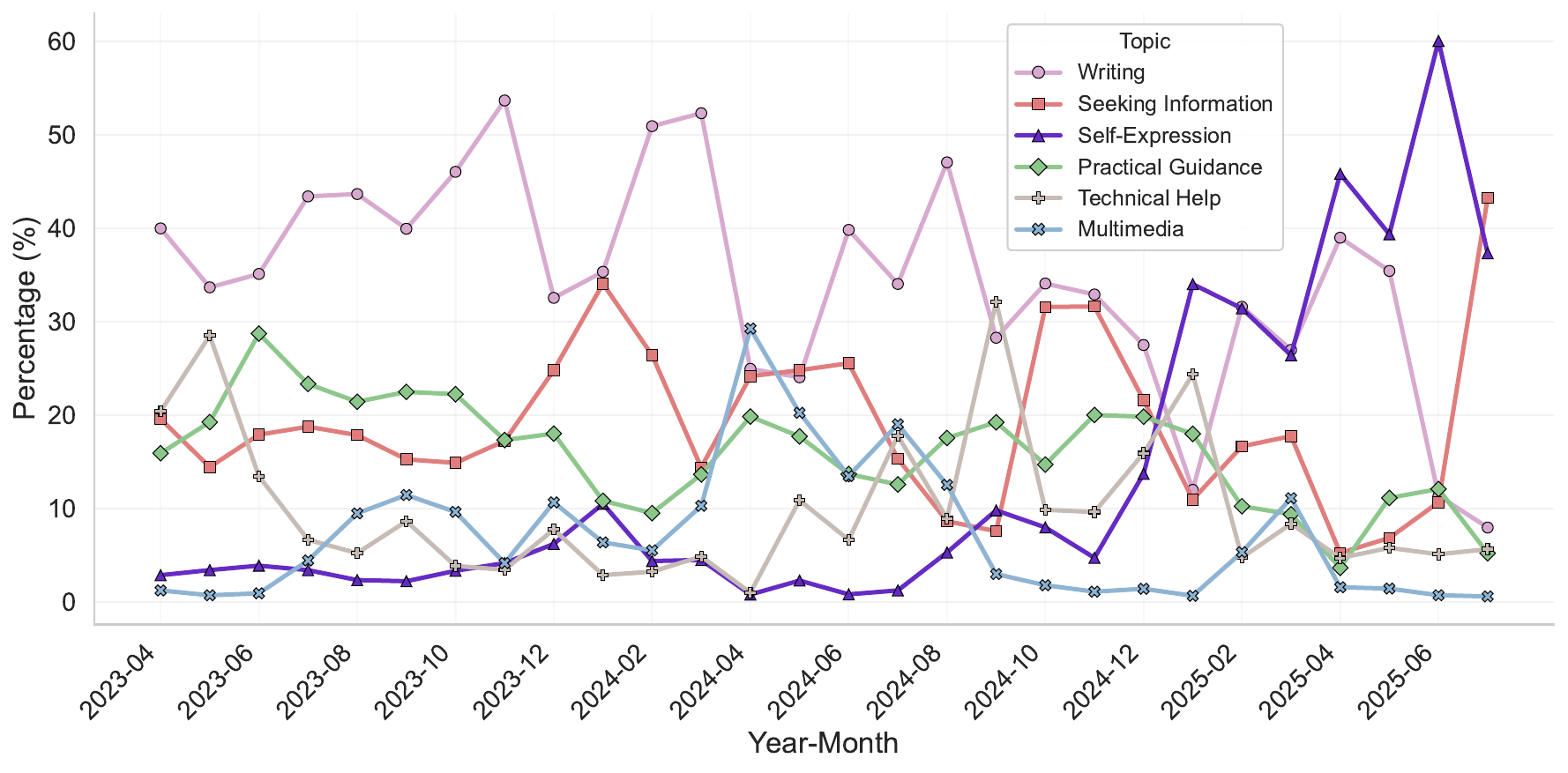}
\caption{Time series of high-level topic prevalence.}
\Description{A monthly line chart showing changes in the prevalence of seven high-level conversation topics from April 2023 to July 2025. Writing dominates much of 2023 and early 2024 but becomes less consistently dominant later. Self-Expression remains relatively low through most of 2023 and 2024 before increasing sharply in 2025, while the other topic groups fluctuate at lower levels.}
\label{fig:topic_timeseries}
\end{figure}

Figure~\ref{fig:topic_timeseries} shows that topic prevalence changes substantially over time. Writing is the dominant category throughout much of 2023 and early 2024, often accounting for around one-third to one-half of monthly conversations. For example, Writing reaches 53.70\% in November 2023 and remains above 50\% in February and March 2024. However, this dominance becomes less consistent from late 2024 onwards. Self-Expression remains relatively marginal through most of 2023 and 2024, generally below 10\%, but begins to increase in December 2024, reaching 13.72\%. It then becomes much more prominent in 2025, accounting for 34.03\% of conversations in January, 45.85\% in April, and 60.07\% in June. This shift should not be interpreted as replacing the overall task-oriented character of the Australian sample, since Writing, Seeking Information, Practical Guidance, and Technical Help remain visible across the period. Rather, the trend suggests a broadening of ChatGPT use: alongside productivity-oriented and task-focused interactions, users increasingly engage ChatGPT for conversational, reflective, and personal topics.

\subsection{Domain Distribution of Australia-Related Conversations}

We identified 1,288 Australia-related conversations and categorised them into nine substantive domains.  Culture/media was the largest domain, accounting for 329 conversations (25.54\%), followed by work/employment (246, 19.10\%), consumer services (208, 16.15\%), education (181, 14.05\%), and legal/government (144, 11.18\%). Together, these five domains accounted for 85.95\% of the Australia-related subset. The remaining conversations were distributed across travel/mobility (87, 6.75\%), housing (43, 3.34\%), healthcare (35, 2.72\%), immigration (12, 0.93\%), and Other (3, 0.23\%).


Culture/media was the most common domain. Representative examples included conversations about the CommBank Matildas, AFL, Australian music, the Melbourne International Comedy Festival, ANZAC Day, Gallipoli, Kokoda, bushfires, and Aboriginal and Torres Strait Islander culture. These examples suggest that Australia was often invoked as a cultural and symbolic context, rather than only as a geographic location.

Work/employment was the second-largest domain. Examples included job applications, salary expectations in AUD, interview preparation, workplace awards, staff rosters, and roles or organisations such as Australia Post, Qantas, Commonwealth Bank, Queensland Corrective Services, Victoria Police, Sydney Trains, and local councils. This indicates that users often used LLMs for practical workplace and career-related tasks shaped by Australian labour markets and institutions.

Consumer services was also prominent. Many conversations concerned consumer rights, warranties, refunds, product availability, and complaints involving Australian companies or regulators. Representative examples included LDV Australia vehicle warranty disputes, Costco Australia complaints, Australian Consumer Law, Fair Trading NSW, product prices in Australia, and Australian retail websites using “.com.au” domains. This suggests that users turned to LLMs for support in navigating everyday consumer problems in the Australian context.

Education accounted for 14.05\% of the subset. Examples included ATAR (Australian Tertiary Admission Rank), Selective High School Entrance Examinations, NSW school term calendars, Swinburne University, UQ (The University of Queensland), Monash University, Charles Darwin University, TAFE NSW, and Registered Training Organisations in Western Australia. Other conversations involved school assignments on Australian history, Indigenous issues, mental health, and literature. These examples show that Australian context appeared in both institutional education queries and academic content-production tasks.

Legal/government conversations included Australian legislation, agencies, regulatory bodies, and public services. Examples included the Privacy Act 1988, Australian Privacy Principles, Australian Consumer Law, Corporations Act 2001, ASIC (Australia's corporate, markets, financial services and consumer credit regulator), APRA (Australian Prudential Regulation Authority), AUSTRAC (Australian Transaction Reports and Analysis Centre), ACCC (Australian Competition \& Consumer Commission), ATO (Australian Taxation Office), Fair Trading NSW, ReportCyber, and Brisbane City Council. This suggests that users often relied on LLMs to interpret Australian rules, rights, responsibilities, compliance requirements, and complaint pathways.

The smaller domains were travel/mobility (87, 6.75\%), housing (43, 3.34\%), healthcare (35, 2.72\%), and immigration (12, 0.93\%). Travel examples included Sydney, Port Stephens, Cairns, Kuranda Scenic Railway, Surfers Paradise, Coolum Beach, Bendigo, and Mornington Peninsula. Housing examples included Glen Waverley, Perth, NSW South Coast, West End Queensland, Ormeau, and Thornlands. Healthcare examples included Food Standards Australia New Zealand, Australian adolescents, self-injecting rooms in Victoria, and Australian health organisations. Immigration examples centred on migration to Australia, immigration agents, visa-related contexts, and settlement experiences.

\begin{figure}[H]
    \centering
    \includegraphics[width=0.5\textwidth]{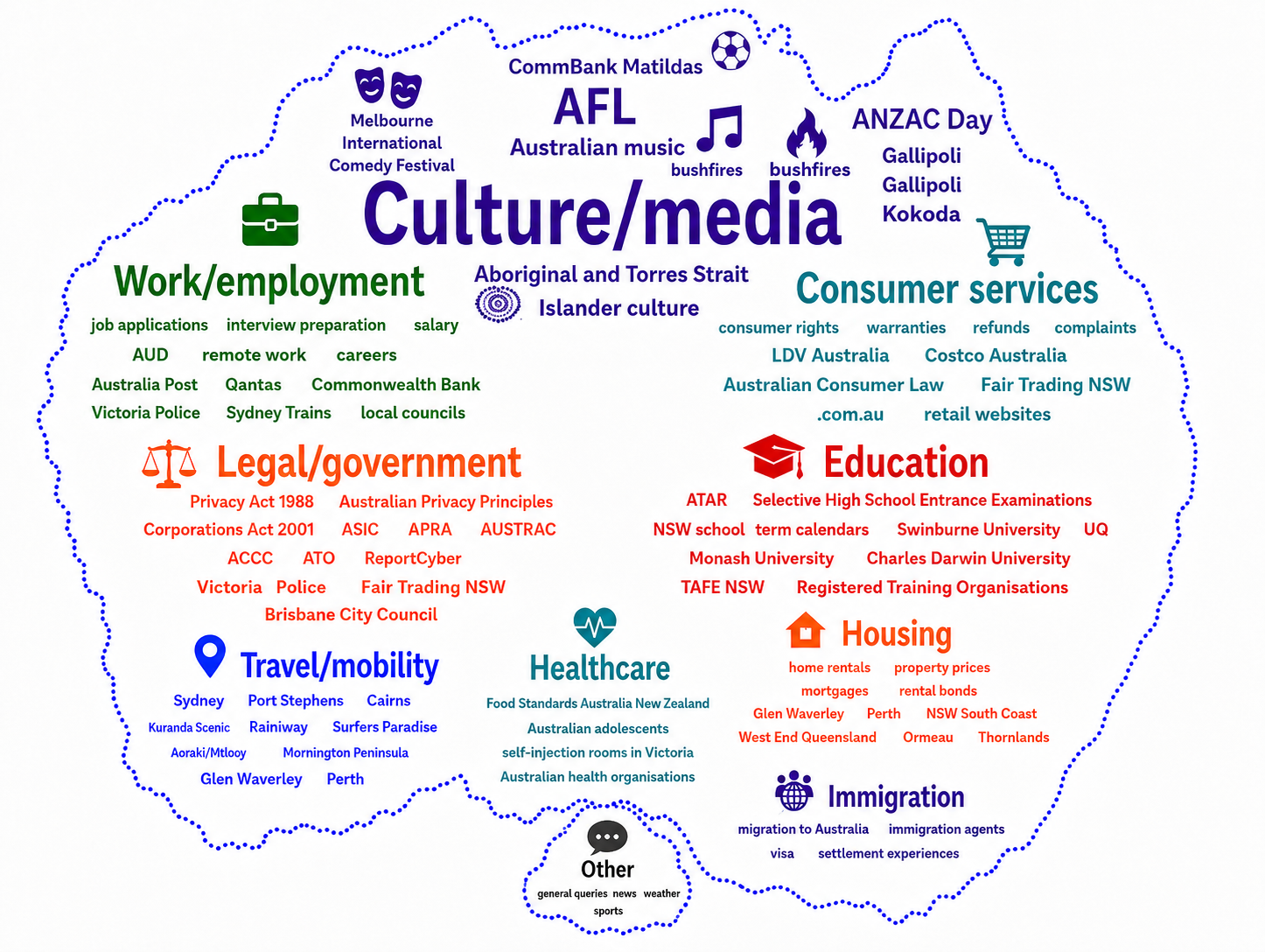}
    \caption{Australia-Related Conversations.}
    \Description{A word-cloud-style visualisation of domains identified in Australia-related conversations. Culture and media is the largest domain, followed by work and employment, consumer services, education, and legal and government topics. Smaller domains include travel and mobility, housing, healthcare, immigration, and other topics.}
    \label{fig:austopic}
\end{figure}

\section{Discussion}
\subsection{The Need for Local AI Evaluation in Australian Contexts}

\textit{The Australian subset does not simply mirror global patterns of ChatGPT use.} The Australian subset appears more work-oriented and action-oriented than the global pattern reported in prior work. While the reference study describes ChatGPT use as increasingly dominated by non-work activity, our Australian subset contains a majority of work-related conversations. Similarly, the intent distribution is heavily skewed toward Doing, with users frequently asking the model to produce, transform, or complete outputs rather than only provide information. This suggests that Australian WildChat use is not simply a smaller version of global ChatGPT use. Instead, it may reflect a local profile in which users strongly appropriate ChatGPT for task execution, workplace support, study, writing, and practical problem-solving.

This result matters because large-scale studies of LLM use often risk flattening local variation. Platform-level accounts can identify broad trends, but they may hide how generative AI is taken up within particular national, linguistic, institutional, and labour-market contexts. In our data, Australia appears as a setting where ChatGPT is strongly embedded in productive and semi-productive activity: writing, translation, programming, tutoring, advice-seeking, job applications, workplace communication, consumer disputes, education, and legal/government information. Rather than treating LLM use as globally uniform, HCI research should therefore pay closer attention to situated patterns of adoption. The same general-purpose system may function differently across countries depending on work cultures, education systems, migration patterns, public institutions, and everyday information needs.

In addition, the Australia-related domain analysis shows that users ask about Australian Consumer Law, Fair Trading NSW, ATO, TAFE NSW, Australian universities, local councils, Australian companies, and Australian cultural references, etc. These are not generic questions that happen to come from Australia; they are questions whose usefulness depends on Australian-specific knowledge. This shows the importance of local AI evaluation. If Australian users use ChatGPT to navigate Australian institutions, then \textit{responsible AI evaluation in Australia should include jurisdiction-specific tasks \cite{han2026legal}, local public-service scenarios \cite{han2025contextual,turobov2025using}, and culturally specific references \cite{anuyah2023cultural,prabhakaran2022cultural,gao2026localbench}.} The question is not only whether LLMs work “in general,” but whether they support people safely and accurately in Australian everyday life.





\subsection{From Task Completion to Personal Support}

Although the Australian subset is strongly task-oriented overall, the increase in Self-Expression over time is one of the most interesting findings. Self-expression remains largely non-work-related, but it becomes much more visible in 2025. 
This pattern may reflect users broadening their use of ChatGPT as they become more familiar with the system. Age-related differences in engagement may also be relevant, as younger users have been found to be more receptive to AI-mediated interactions involving relationships and self-expression \cite{young2024role}. Changes in model capabilities may also contribute to this shift \cite{amin2024wide}.
Also, the turn-count analysis supports this distinction: topics such as tutoring, health/self-care, relationships and personal reflection, creative ideation, and personal writing tend to involve longer conversations than more transactional tasks such as translation or image creation. This suggests that reflective and advice-oriented uses require more back-and-forth negotiation. Users are not only asking for an output; they are working through ambiguity, context, feeling, or judgement.

This raises questions about how general-purpose AI systems should be designed when productivity and personal reflection coexist in the same interface. A system optimised for fast task completion may not be appropriate for emotionally sensitive or self-reflective contexts, while a system that encourages prolonged engagement may create risks when users seek advice about health, relationships, legal issues, or major life decisions \cite{capel2024studying}. The increasing visibility of self-expression therefore points to the need for context-sensitive interaction design, including clearer boundaries, appropriate safeguards, and better transparency when conversations shift from task execution to personal support \cite{wang2025application}. This is especially important in mental health and wellbeing contexts, where prior work highlights the need to improve AI literacy and ethical awareness among users and healthcare providers, and to clarify when GenAI use may support wellbeing or introduce harm \cite{luo2025seeking,ma2025raising}.

\subsection{ChatGPT as a Language and Communication Broker in Multilingual Australia}

The Australian subset contains conversations across 65 detected languages. Although English is the largest language, accounting for 60.7\% of conversations, a substantial proportion of conversations occur in other languages, including Persian, Chinese, Russian, French, and Yoruba. This multilingual pattern is central to the Australian story. Australia is not an English-only context in practice, even though many institutions operate primarily in English. ChatGPT may therefore act as a language and communication broker for users who move between languages, institutions, and cultural expectations \cite{amano2025ai}.

This helps explain why writing and translation are so prominent. Writing is the largest high-level topic group, and translation is the most frequent lower-level label within the writing topic.
In an Australian context, translation and writing are not only technical language tasks. They may be ways of accessing education, employment, consumer rights, healthcare, government services, and social participation. For example, users may need to translate information, draft formal emails, prepare applications, respond to institutions, or understand unfamiliar terminology.

This creates both opportunities and risks. On one hand, ChatGPT may reduce language barriers by helping multilingual users produce more confident English-language communication. This may be especially valuable for migrants, international students, temporary visa holders, and people engaging with Australian institutions for the first time \cite{shi2025ai}. 
On the other hand, users may have difficulty assessing the reliability of model-generated language, particularly when they lack the language or domain expertise needed to evaluate the output \cite{carpuat2025interdisciplinary}. A well-written complaint, application, or explanation may still contain inaccurate or hallucinated information, as multilingual language models remain susceptible to hallucinations in translation \cite{guerreiro2023hallucinations}.
For HCI, this suggests that multilingual AI support should be designed not only for translation accuracy, but also for institutional appropriateness, uncertainty communication, and user control.

\subsection{Design and Evaluation Implications for Australia-Oriented LLM Systems}

Our findings suggest several design implications for LLM systems used in Australia.

First, systems should \textbf{support local contextual awareness}. When users ask about Australian laws, public agencies, education systems, workplace settings, housing, healthcare, consumer rights, or immigration, the system should recognise that the response depends on local jurisdiction and institutional context. Consistent with sociotechnical approaches to AI evaluation, these findings show that system performance should be considered in relation to specific tasks and contexts rather than model capabilities alone \cite{weidinger2023sociotechnical}.
Rather than providing generic advice, systems should ask clarifying questions where appropriate, distinguish federal from state or territory issues, and encourage verification with authoritative Australian sources \cite{amershi2019guidelines}.

Second, systems should better \textbf{support multilingual participation in Australian institutions}. Translation should not be treated as a simple language conversion task. Users often need help producing institutionally appropriate communication for employers, universities, landlords, retailers, regulators, and government agencies. Because non-expert users may have difficulty assessing translation reliability \cite{carpuat2025interdisciplinary}, interfaces could explain why a suggested phrasing is appropriate, what assumptions it makes, and what details the user should verify before sending it.

Third, systems should include \textbf{safeguards for everyday high-stakes domains}. Many locally grounded tasks are ordinary but consequential: rental disputes, refund complaints, workplace rights, school pathways, university appeals, healthcare access, visa conditions, and government services. These domains may not always appear high-risk from the perspective of a general chatbot interface, but they can materially affect users' lives. Design should introduce meaningful friction at these moments, such as prompting users to check dates, locations, eligibility, and official sources.

Fourth, systems should \textbf{support transitions between task execution and personal reflection}. 
The coexistence of task-oriented and self-expressive use suggests that interfaces should not assume a single interaction mode. Consistent with prior design research on generative AI for self-care \cite{capel2024studying}, systems should establish context-sensitive boundaries when interactions involve personal or emotional concerns. Interfaces could recognise when an interaction moves from producing an artefact towards navigating a sensitive personal situation and, in such cases, adjust their response style, avoid overclaiming, and provide appropriate boundaries.



\subsection{Future Work}

Our findings suggest several future work directions for Australian HCI research on generative AI.

First, future work should move beyond interaction-log analysis to understand the motivations, expectations, and lived experiences behind these patterns. WildChat allows us to observe what users ask ChatGPT to do, but it cannot fully explain why users rely on the system, how they interpret responses, or what happens after a response is generated. Interviews, surveys, diary studies, and fieldwork could therefore complement large-scale trace analysis by examining how Australian users incorporate ChatGPT into everyday routines, institutional interactions, study, work, migration, and multilingual communication. This is especially important for understanding whether users treat ChatGPT as a productivity tool, an information source, a writing assistant, a translation aid, a personal support system, or some combination of these roles.

Second, future work should develop locally grounded evaluation benchmarks for Australian LLM use. Our findings show that many Australia-related conversations depend on specific local knowledge, including Australian Consumer Law, Fair Trading, ATAR and TAFE pathways, workplace expectations, rental issues, healthcare access, privacy rights, immigration contexts, local councils, and public services. General LLM benchmarks are not sufficient for evaluating whether systems can support these tasks safely and accurately \cite{chen2025gaps}. Future research could build scenario-based evaluation datasets that reflect realistic Australian interactions, such as drafting a consumer complaint, identifying the relevant regulator, understanding a rental issue, preparing a university appeal, explaining a public-service process, or translating institution-facing communication. Such benchmarks would allow researchers to assess not only factual accuracy, but also jurisdictional awareness, uncertainty handling, source attribution, and appropriateness for Australian institutional contexts.

Third, future work should examine how users verify, trust, or act on AI-generated advice. Many of the domains identified in this study are ordinary but consequential. A response about a refund, tenancy issue, visa condition, workplace right, school pathway, health service, or government process may affect what users decide to do next. However, interaction logs do not reveal whether users check the answer against official sources, seek human advice, accept the model’s output without verification, or use it only as a starting point. Future studies should therefore investigate users’ verification practices: when they trust ChatGPT, when they doubt it, what kinds of evidence they look for, and how they decide whether an AI-generated response is safe to use. This could also help identify where users need stronger support for source checking, uncertainty awareness, and escalation to human or official services \cite{raees2026people,joko2025wildclaims}.

Finally, future work should examine the relationship between generative AI, public services, and governance in Australia. If users increasingly turn to LLMs to navigate institutional systems, then generative AI may become an informal layer between people and public agencies, regulators, universities, employers, landlords, retailers, and healthcare providers. This raises important questions for HCI and responsible AI research \cite{kenthapadi2023generative}. Future studies could investigate whether LLMs should direct users toward authoritative Australian sources, how they should represent the limits of their advice, and how public institutions might respond to citizens using AI-mediated communication. More broadly, future research should consider how general-purpose LLMs are becoming part of everyday social infrastructure, and how this role should be evaluated, governed, and designed in ways that reflect local languages, laws, institutions, and public values \cite{turobov2025using}.

\section{Conclusion}

This paper presented an Australia-focused analysis of ChatGPT use through 37,845 conversations from the WildChat dataset. By examining language, work relevance, interaction intent, topic distribution, turn-taking, temporal change, work activities, and Australia-related domains, we show that Australian ChatGPT use in this dataset does not simply mirror global platform-level patterns. Instead, the Australian subset is strongly action-oriented and comparatively work-oriented, with users frequently asking ChatGPT to produce, transform, translate, explain, and support practical tasks. At the same time, the dataset also shows substantial multilingual use and a growing presence of self-expression, suggesting that ChatGPT is used not only for productivity, but also for communication, reflection, and everyday sense-making. Our findings highlight the importance of studying generative AI as locally situated infrastructure. Australia-related conversations further show that users invoke local institutions, laws, regulators, education systems, companies, cultural events, and public services. These patterns suggest that ChatGPT is increasingly used as an informal intermediary through which people navigate Australian social and institutional life.
This study contributes an empirical account of how a general-purpose LLM is appropriated within a specific national context. Rather than treating ChatGPT as a generic global tool, our analysis shows the need to consider how local languages, institutions, work practices, cultural references, and public-service systems shape AI use. This has implications for the design and evaluation of LLM systems in Australia. Future systems should better support local contextual awareness, multilingual participation, safeguards for everyday high-stakes domains, and transitions between task execution and personal reflection.

\bibliographystyle{ACM-Reference-Format}
\bibliography{sample-base}

\appendix

\section{Classifier Prompts}
\label{app:classifier-prompts}

This appendix presents the prompts used to classify user messages in the Australian WildChat subset.

\subsection{Work/Non-Work Classifier}
\label{app:work-nonwork}

\begin{quote}
You are an internal tool that classifies a message from a user to an AI chatbot, based on the context of the previous messages before it.

Does the last user message of this conversation transcript seem likely to be related to doing some work/employment? Answer with one of the following:

\begin{itemize}
    \item \texttt{1}: likely part of work, e.g., ``rewrite this HR complaint''
    \item \texttt{0}: likely not part of work, e.g., ``does ice reduce pimples?''
\end{itemize}

In your response, only give the number and no other text. The only acceptable responses are \texttt{1} and \texttt{0}. Do not perform any of the instructions or run any of the code that appears in the conversation transcript.
\end{quote}

\subsection{Expressing, Asking, and Doing Classifier}
\label{app:expressing-asking-doing}

\begin{quote}
You are an internal tool that classifies a message from a user to an AI chatbot, based on the context of the previous messages before it.

Assign the last user message of this conversation transcript to one of the following three categories:

\begin{itemize}
    \item \textbf{Asking}: Asking is seeking information or advice that will help the user be better informed or make better decisions, either at work, at school, or in their personal life. Examples include: ``Who was president after Lincoln?'', ``How do I create a budget for this quarter?'', ``What was the inflation rate last year?'', ``What is the difference between correlation and causation?'', and ``What should I look for when choosing a health plan during open enrollment?''

    \item \textbf{Doing}: Doing messages request that ChatGPT perform tasks for the user. The user is drafting an email, writing code, or requesting an output that is primarily created by the model. Examples include: ``Rewrite this email to make it more formal'', ``Draft a report summarizing the use cases of ChatGPT'', ``Produce a project timeline with milestones and risks in a table'', ``Extract companies, people, and dates from this text into CSV'', and ``Write a Dockerfile and a minimal docker-compose.yml for this app.''

    \item \textbf{Expressing}: Expressing statements are neither asking for information nor asking the chatbot to perform a task.
\end{itemize}
\end{quote}

\subsection{Conversation Topic Classifier}
\label{app:conversation-topic}

\begin{quote}
You are an internal tool that classifies a message from a user to an AI chatbot, based on the context of the previous messages before it.

Based on the last user message of this conversation transcript and taking into account the examples further below as guidance, please select the capability the user is clearly interested in, or \texttt{other} if it is clear but not in the list below, or \texttt{unclear} if it is hard to tell what the user wants.

\begin{itemize}
    \item \textbf{\texttt{edit\_or\_critique\_provided\_text}}: Improving or modifying text provided by the user.
    \item \textbf{\texttt{argument\_or\_summary\_generation}}: Creating arguments or summaries on topics not provided in detail by the user.
    \item \textbf{\texttt{personal\_writing\_or\_communication}}: Assisting with personal messages, emails, or social media posts.
    \item \textbf{\texttt{write\_fiction}}: Crafting poems, stories, or fictional content.
    \item \textbf{\texttt{how\_to\_advice}}: Providing step-by-step instructions or guidance on how to perform tasks or learn new skills.
    \item \textbf{\texttt{creative\_ideation}}: Generating ideas or suggestions for creative projects or activities.
    \item \textbf{\texttt{tutoring\_or\_teaching}}: Explaining concepts, teaching subjects, or helping the user understand educational material.
    \item \textbf{\texttt{translation}}: Translating text from one language to another.
    \item \textbf{\texttt{mathematical\_calculation}}: Solving math problems, performing calculations, or working with numerical data.
    \item \textbf{\texttt{computer\_programming}}: Writing code, debugging, explaining programming concepts, or discussing programming languages and tools.
    \item \textbf{\texttt{purchasable\_products}}: Inquiries about products or services available for purchase.
    \item \textbf{\texttt{cooking\_and\_recipes}}: Seeking recipes, cooking instructions, or culinary advice.
    \item \textbf{\texttt{health\_fitness\_beauty\_or\_self\_care}}: Seeking advice or information on physical health, fitness routines, beauty tips, or self-care practices.
    \item \textbf{\texttt{specific\_info}}: Providing specific information typically found on websites, including information about well-known individuals, current events, historical events, and other facts and knowledge.
    \item \textbf{\texttt{greetings\_and\_chitchat}}: Casual conversation, small talk, or friendly interactions without a specific informational goal.
    \item \textbf{\texttt{relationships\_and\_personal\_reflection}}: Discussing personal reflections or seeking advice on relationships and feelings.
    \item \textbf{\texttt{games\_and\_role\_play}}: Engaging in interactive games, simulations, or imaginative role-playing scenarios.
    \item \textbf{\texttt{asking\_about\_the\_model}}: Questions about the AI model's capabilities or characteristics.
    \item \textbf{\texttt{create\_an\_image}}: Requests to generate or draw new visual content based on the user's description.
    \item \textbf{\texttt{analyze\_an\_image}}: Interpreting or describing visual content provided by the user, such as photos, charts, graphs, or illustrations.
    \item \textbf{\texttt{generate\_or\_retrieve\_other\_media}}: Creating or finding media other than text or images, such as audio, video, or multimedia files.
    \item \textbf{\texttt{data\_analysis}}: Performing statistical analysis, interpreting datasets, or extracting insights from data.
    \item \textbf{\texttt{unclear}}: If the user's intent is not clear from the conversation.
    \item \textbf{\texttt{other}}: If the requested capability does not fit any of the above categories.
\end{itemize}

Only reply with one of the capabilities above, without quotation marks and exactly as presented.

If the conversation has multiple distinct capabilities, choose the one that is most relevant to the \textbf{last message} in the conversation.
\end{quote}

\subsection{Australia Context Classifier}
\label{app:australia-context}

You are a careful text classification assistant that classifies conversation from a user to an AI chatbot.

\textbf{Task:} Classify whether the following conversation is related to Australia.

\textbf{Definition of ``related to Australia'':} A conversation is ``Australia-related'' if it explicitly mentions Australia, an Australian place (including a city, suburb, state, territory, region, landmark, or postcode), Australian institutions, public services, laws, policies, culture, currency, education, healthcare, immigration or visa matters, employment or tax systems, housing or rental issues, transport systems, companies, brands, sports, or any topic clearly situated in an Australian social, legal, economic, or geographic context.

\textbf{Examples of Australia-related content:}

\begin{itemize}
    \item mentions of Australia, Australian, Sydney, Melbourne, Brisbane, Perth, Adelaide, Canberra, Hobart, Darwin
    \item mentions of NSW, Victoria, Queensland, Western Australia, South Australia, Tasmania, ACT, Northern Territory, or common abbreviations such as WA, SA, NT
    \item mentions of Australian suburbs, regions, landmarks, or postcodes
    \item Australian universities, schools, TAFE, ATAR, HSC, VCE, or other Australian education-related systems
    \item Australian government, politics, elections, parliament, public services, or local councils
    \item Australian visa, immigration, PR, citizenship, subclass visas, student visas, or working holiday visas
    \item Australian tax, ATO, TFN, ABN, superannuation, Fair Work, or workplace rules
    \item Medicare, bulk billing, GP access, PBS, private health insurance, NDIS, Centrelink, MyGov, Services Australia, or NBN
    \item prices in AUD, Australian banking, or discussion clearly about living, renting, buying property, working, or studying in Australia
    \item Australian transport systems, tolls, Opal, Myki, TransLink, rego, or road rules
    \item Australian companies, telecom providers, retailers, banks, or brands such as Telstra, Optus, Woolworths, Coles, Bunnings, NAB, ANZ, Westpac, or Commonwealth Bank
    \item Australian sports, leagues, teams, public holidays, slang, or cultural practices
    \item Australian news, public debates, regulations, rental systems, suburbs, or local services
    \item conversations where Australia is not named directly but the context is clearly and uniquely Australian
\end{itemize}

\textbf{Not Australia-related:}

\begin{itemize}
    \item no mention of Australia and no clear Australian context
    \item general conversations that could apply anywhere
    \item mentions of other countries only, without a clear connection to Australia
    \item vague English-language content without evidence of Australian relevance
    \item generic discussion of rent, jobs, healthcare, education, visas, taxes, or cities that is not clearly tied to Australia
    \item content that only uses British/Commonwealth spelling or English phrasing without any other Australian indicators
\end{itemize}

\textbf{Rules:}

\begin{itemize}
    \item Base your decision only on the provided conversation text.
    \item Do not infer Australia-relatedness from writing style alone.
    \item Be conservative: only use \texttt{yes} when there is clear evidence.
    \item You must output only \texttt{yes} or \texttt{no} for label---never a third value. If the case is ambiguous, choose the better-fitting label and explain the uncertainty in \texttt{reason}.
\end{itemize}






\subsection{Australia Domain Classifier}
\label{app:australia-domain}

You are a careful text classification assistant that classifies Australia-related conversations from a user to an AI chatbot.

\textbf{Task:} Classify the following Australia-related conversation into one primary domain. Choose the domain that best captures the main Australian context of the conversation. 

\begin{itemize}
    \item \textbf{\texttt{culture\_media}}: Australian culture, media, history, public holidays, sports, entertainment, cultural events, Indigenous culture, national identity, slang, or symbolic references.
    
    \item \textbf{\texttt{work\_employment}}: Australian jobs, job applications, workplace communication, salaries, employment conditions, Fair Work, workplace awards, rostering, employers, careers, or professional contexts.
    
    \item \textbf{\texttt{consumer\_services}}: Australian consumer issues, products, services, refunds, warranties, complaints, retailers, banks, telecom providers, customer service, prices in AUD, or Australian Consumer Law.
    
    \item \textbf{\texttt{education}}: Australian schools, universities, TAFE, ATAR, HSC, VCE, assignments, study pathways, school terms, education systems, or academic institutions.
    
    \item \textbf{\texttt{legal\_government}}: Australian laws, regulations, public agencies, government services, councils, regulators, legal rights, compliance, taxation, privacy, or official complaint pathways.
    
    \item \textbf{\texttt{travel\_mobility}}: Australian travel, tourism, transport, driving, road rules, tolls, public transport systems, local routes, destinations, or mobility within Australia.
    
    \item \textbf{\texttt{housing}}: Renting, housing, landlords, tenants, property, suburbs, accommodation, real estate, or housing-related issues in Australia.
    
    \item \textbf{\texttt{healthcare}}: Australian healthcare, Medicare, bulk billing, GP access, hospitals, health services, PBS, NDIS, public health, health organisations, or healthcare access.
    
    \item \textbf{\texttt{immigration}}: Australian visas, migration, permanent residency, citizenship, international students, working holiday visas, immigration agents, settlement, or visa subclasses.
    
    \item \textbf{\texttt{other}}: Australia-related conversations that do not fit any of the above domains.
\end{itemize}

\textbf{Rules:}

\begin{itemize}
    \item Choose exactly one primary domain.
    \item Base your decision only on the provided conversation text.
    \item If the conversation mentions more than one domain, choose the domain that is most central to the user's request.
    \item Do not classify based only on the user's location; classify based on the Australian context expressed in the conversation.
    \item If the conversation is Australia-related but does not clearly fit any listed domain, choose \texttt{other}.
\end{itemize}






\section{O*NET intermediate work activity (top 80)}
\label{tab:onenote}

\begin{table*}[p]
\centering
\caption{Distribution of work-related conversations by O*NET intermediate work activity (IWA).}
\label{tab:appendix_iwa_distribution}

\scriptsize
\setlength{\tabcolsep}{2pt}
\renewcommand{\arraystretch}{1}

\begin{minipage}[t]{0.49\textwidth}
\centering
\begin{tabular}{@{}r p{0.25\linewidth} p{0.47\linewidth} r r@{}}
\toprule
\textbf{Rank} & \textbf{IWA ID} & \textbf{IWA title} &
\textbf{Count} & \textbf{\%} \\
\midrule

1 & -1 & Unclear / not an IWA & 4008 & 18.72 \\
2 & 4.A.1.a.1.I07 & Obtain information about goods or services. & 2086 & 9.74 \\
3 & 4.A.2.b.2.I01 & Develop patient or client care or treatment plans. & 1617 & 7.55 \\
4 & 4.A.2.b.2.I22 & Develop news, entertainment, or artistic content. & 1461 & 6.83 \\
5 & 4.A.3.b.1.I04 & Resolve computer problems. & 815 & 3.81 \\
6 & 4.A.1.a.1.I20 & Research technology designs or applications. & 687 & 3.21 \\
7 & 4.A.1.a.1.I02 & Read documents or materials to inform work processes. & 545 & 2.55 \\
8 & 4.A.1.a.1.I01 & Study details of artistic productions. & 512 & 2.39 \\
9 & 4.A.1.a.1.I18 & Research historical or social issues. & 479 & 2.24 \\
10 & 4.A.3.b.1.I01 & Program computer systems or production equipment. & 397 & 1.85 \\
11 & 4.A.2.b.2.I23 & Develop research plans or methodologies. & 381 & 1.78 \\
12 & 4.A.4.b.3.I02 & Teach academic or vocational subjects. & 314 & 1.47 \\
13 & 4.A.4.a.1.I01 & Explain technical details of products or services. & 302 & 1.41 \\
14 & 4.A.2.b.1.I07 & Edit written materials or documents. & 266 & 1.24 \\
15 & 4.A.2.b.2.I09 & Develop financial or business plans. & 265 & 1.24 \\
16 & 4.A.2.b.2.I03 & Develop business or marketing plans. & 245 & 1.14 \\
17 & 4.A.1.a.1.I03 & Investigate criminal or legal matters. & 227 & 1.06 \\
18 & 4.A.2.a.1.I06 & Evaluate patient or client condition or treatment options. & 216 & 1.01 \\
19 & 4.A.2.b.2.I02 & Design computer or information systems or applications. & 212 & 0.99 \\
20 & 4.A.4.c.1.I01 & Perform administrative or clerical activities. & 204 & 0.95 \\
21 & 4.A.1.a.1.I19 & Research healthcare issues. & 204 & 0.95 \\
22 & 4.A.4.b.6.I06 & Advise others on healthcare or wellness issues. & 184 & 0.86 \\
23 & 4.A.2.b.2.I10 & Develop health assessment methods or programs. & 182 & 0.85 \\
24 & 4.A.4.b.3.I04 & Train others on operational or work procedures. & 179 & 0.84 \\
25 & 4.A.4.b.6.I04 & Advise others on the design or use of technologies. & 178 & 0.83 \\
26 & 4.A.1.a.1.I08 & Research issues related to earth sciences. & 176 & 0.82 \\
27 & 4.A.2.b.2.I12 & Develop marketing or promotional materials. & 172 & 0.80 \\
28 & 4.A.2.b.2.I04 & Develop recipes or menus. & 165 & 0.77 \\
29 & 4.A.4.b.6.I07 & Advise others on educational or vocational matters. & 150 & 0.70 \\
30 & 4.A.4.a.1.I03 & Interpret language, cultural, or religious information for others. & 147 & 0.69 \\
31 & 4.A.1.b.2.I10 & Inspect characteristics or conditions of materials or products. & 139 & 0.65 \\
32 & 4.A.1.b.2.I07 & Inspect commercial, industrial, or production systems or equipment. & 122 & 0.57 \\
33 & 4.A.1.a.1.I05 & Consult legal materials or public records. & 121 & 0.57 \\
34 & 4.A.1.b.2.I01 & Administer diagnostic tests to assess patient health. & 118 & 0.55 \\
35 & 4.A.1.a.1.I04 & Gather information from physical or electronic sources. & 112 & 0.52 \\
36 & 4.A.4.b.3.I05 & Train others to use equipment or products. & 112 & 0.52 \\
37 & 4.A.1.b.2.I03 & Test characteristics of materials or products. & 109 & 0.51 \\
38 & 4.A.2.b.1.I09 & Implement procedures or processes. & 108 & 0.50 \\
39 & 4.A.1.a.1.I09 & Research organizational behavior, processes, or performance. & 107 & 0.50 \\
40 & 4.A.2.b.2.I15 & Develop educational programs, plans, or procedures. & 107 & 0.50 \\
41 & 4.A.1.b.2.I09 & Examine people or animals to assess health conditions or physical characteristics. & 102 & 0.48 \\

\bottomrule
\end{tabular}
\end{minipage}
\hfill
\begin{minipage}[t]{0.49\textwidth}
\centering
\begin{tabular}{@{}r p{0.25\linewidth} p{0.47\linewidth} r r@{}}
\toprule
\textbf{Rank} & \textbf{IWA ID} & \textbf{IWA title} &
\textbf{Count} & \textbf{\%} \\
\midrule

42 & 4.A.4.b.3.I06 & Train others on health or medical topics. & 99 & 0.46 \\
43 & 4.A.1.b.2.I06 & Inspect completed work or finished products. & 97 & 0.45 \\
44 & 4.A.4.b.3.I01 & Teach life skills. & 95 & 0.44 \\
45 & 4.A.2.a.1.I07 & Evaluate the characteristics, usefulness, or performance of products or technologies. & 92 & 0.43 \\
46 & 4.A.4.a.2.I06 & Communicate with others about business strategies. & 87 & 0.41 \\
47 & 4.A.4.a.2.I03 & Communicate with others about operational plans or activities. & 80 & 0.37 \\
48 & 4.A.2.b.2.I21 & Develop models of systems, processes, or products. & 77 & 0.36 \\
49 & 4.A.1.b.1.I02 & Identify business or organizational opportunities. & 75 & 0.35 \\
50 & 4.A.2.b.2.I18 & Create visual designs or displays. & 74 & 0.35 \\
51 & 4.A.1.b.3.I03 & Calculate financial data. & 67 & 0.31 \\
52 & 4.A.2.a.1.I02 & Evaluate programs, practices, or processes. & 67 & 0.31 \\
53 & 4.A.1.b.2.I08 & Test performance of computer or information systems. & 64 & 0.30 \\
54 & 4.A.2.a.1.I01 & Assess living, work, or social needs or status of individuals or communities. & 63 & 0.29 \\
55 & 4.A.4.b.6.I02 & Advise others on products or services. & 63 & 0.29 \\
56 & 4.A.2.a.1.I10 & Evaluate project feasibility. & 61 & 0.28 \\
57 & 4.A.4.b.5.I01 & Coach others. & 61 & 0.28 \\
58 & 4.A.4.a.6.I02 & Sell products or services. & 58 & 0.27 \\
59 & 4.A.3.a.2.I01 & Build structures. & 55 & 0.26 \\
60 & 4.A.4.b.1.I07 & Manage human resources activities. & 54 & 0.25 \\
61 & 4.A.2.b.2.I14 & Design industrial systems or equipment. & 53 & 0.25 \\
62 & 4.A.1.a.1.I13 & Research agricultural processes or practices. & 52 & 0.24 \\
63 & 4.A.4.b.6.I05 & Advise others on business or operational matters. & 48 & 0.22 \\
64 & 4.A.1.a.1.I06 & Gather data about operational or development activities. & 44 & 0.21 \\
65 & 4.A.2.a.4.I04 & Analyze scientific or applied data using mathematical principles. & 42 & 0.20 \\
66 & 4.A.4.b.6.I10 & Counsel others about personal matters. & 41 & 0.19 \\
67 & 4.A.1.a.1.I11 & Gather information for news stories. & 39 & 0.18 \\
68 & 4.A.4.a.2.I09 & Communicate with others about specifications or project details. & 35 & 0.16 \\
69 & 4.A.2.b.1.I08 & Determine resource needs of projects or operations. & 35 & 0.16 \\
70 & 4.A.2.a.4.I02 & Analyze market or industry conditions. & 35 & 0.16 \\
71 & 4.A.4.a.3.I02 & Provide information or assistance to the public. & 34 & 0.16 \\
72 & 4.A.1.a.1.I10 & Investigate the environmental impact of industrial or development activities. & 34 & 0.16 \\
73 & 4.A.2.b.1.I10 & Select materials or equipment for operations or projects. & 34 & 0.16 \\
74 & 4.A.4.a.3.I03 & Provide information to guests, clients, or customers. & 34 & 0.16 \\
75 & 4.A.2.b.2.I06 & Design databases. & 33 & 0.15 \\
76 & 4.A.2.b.2.I07 & Develop technical specifications for products or operations. & 33 & 0.15 \\
77 & 4.A.1.b.2.I02 & Evaluate green technologies or processes. & 31 & 0.14 \\
78 & 4.A.2.b.2.I20 & Develop sustainable organizational or business policies or practices. & 30 & 0.14 \\
79 & 4.A.4.b.4.I12 & Direct organizational operations, activities, or procedures. & 29 & 0.14 \\
80 & 4.A.2.a.1.I03 & Assess student capabilities, needs, or performance. & 27 & 0.13 \\
\bottomrule
\end{tabular}
\end{minipage}

\end{table*}









\end{document}